\documentclass[a4paper, journal, 10pt]{IEEEtran}
\normalsize

\usepackage{cite}
\usepackage{bm}
\usepackage{amsmath}
\usepackage{amsthm}
\newtheorem{defn}{Definition} 
\usepackage{amssymb}

\usepackage{caption}
\usepackage{subcaption}

\newcommand{\LRT}[2]{%
	\mathrel{\mathop\gtrless\limits^{#1}_{#2}}%
}

\usepackage[final]{graphicx}
\begin{document}

\title{Improved Detection Performance of Passive Radars Exploiting Known Communication Signal Form}
%
%
%

\author{ Anantha K. Karthik, \IEEEmembership{Student Member, IEEE} and Rick S. Blum, \IEEEmembership{IEEE Fellow} 
        
\thanks{This material is based upon work partially supported by the U. S. Army Research Laboratory and the U. S. Army Research Office under grant number W911NF-17-1-0331 and by the National Science Foundation under grant no. ECCS-1405579.}

\thanks{Anantha K. Karthik and Rick S. Blum are with the Department of Electrical and Computer Engineering, Lehigh University, Bethlehem, PA 18015 USA (e-mail: akk314@lehigh.edu; rblum@lehigh.edu).}}

\maketitle

\begin{abstract}	
In this paper, we address the problem of target detection in passive multiple-input-multiple-output (MIMO) radar networks. A generalized likelihood ratio test is derived, assuming prior knowledge of the signal format used in the  non-cooperative transmit stations.  We consider scenarios in which the unknown transmitted signal uses either a linear digital modulation scheme or the Orthogonal Frequency-Division Multiplexing (OFDM) modulation scheme. These digital modulation schemes are used in popular standards including Code-Division Multiple Access (CDMA), Digital Video Broadcasting-Terrestrial (DVB-T) and Long Term Evaluation (LTE). The performance of the generalized likelihood ratio test in the known signal format case is often significantly more favorable when compared to the case that does not exploit this information. Further, the performance improves with increasing number of samples per symbol and, for a sufficiently large number of samples per symbol, the performance closely approximates that of an active radar with a known transmitted signal.
\end{abstract}


%
\IEEEpeerreviewmaketitle

\section{Introduction}\label{section1}
Passive radar differs from conventional active radar in that it relies on preexisting signals from non-cooperative transmitters instead of transmitting a known signal. Examples of non-cooperative transmitters include radio transmitters, TV transmitters, cellular base stations, and other such high power transmitters. Such a system is cost efficient, covert, and suitable for emergencies due to the lack of a transmitter. On the other hand, removing the transmitter typically adds significant complexity to the signal processing algorithms needed in the system.

Consider a scenario where the passive radar system utilizes the signals transmitted from a cellular base station for target detection. Although we do not control the base station, we usually have prior information regarding the position of the transmitter and signal format used in the base station. However, the transmitted signal still contains unknown information bits, so the signal is not fully known. Prior publications available in the literature derived explicit closed-form expressions for the Generalized Likelihood Ratio Tests (GLRTs) for target detection in Passive MIMO Radar (PMR) networks \cite{Hack_PMR, Hack_PMR_conf, Hack_noRef, Liu, Cui, Wang}. However, they did not consider the possibility of exploiting the available signal format information.

The authors in \cite{Hack_PMR, Hack_PMR_conf, Hack_noRef, Liu} derived the GLRT for target detection in PMR networks depending on whether the direct-path reference channel signal is available \cite{Hack_PMR, Hack_PMR_conf} or not \cite{Hack_noRef, Liu}. In \cite{Hack_PMR, Hack_PMR_conf, Hack_noRef, Liu}, the discrete-time samples of the transmitted signal are assumed to be a deterministic unknown parameter. The transmitted signal along with other unknown parameters are estimated in the GLRT procedure. This work in \cite{Hack_PMR, Hack_PMR_conf, Hack_noRef, Liu} appears to be the closest previous work to that discussed in the current paper. Furthermore, the authors in \cite{Liu} derived the GLRT under conditions where the noise variance is either known or unknown.

In \cite{Cui, Wang, Zhang}, the authors derived other GLRTs for target detection in scenarios where the unknown transmitted signal is stochastic. A circular Gaussian random variable with zero mean and unit variance is used to model the transmitted signal. A related paper \cite{Zhang} derived the GLRT for PMR networks in which the direct-path and reflected-path signals are not separated. The unknown parameters necessary for the GLRT are estimated using the Expectation-Maximization (EM) algorithm. Target estimation has also been considered for passive radar, see \cite{QianHe2010, Stinco, QianHe2014}.

In this paper, we study the problem of target detection in PMR networks assuming prior knowledge of the signal format of the transmitted signal. The known signal format scenario is not unrealistic as many cellular base stations or other transmit stations emit signals according to known standard protocols. In our work, we consider scenarios in which the transmitted signal uses either a linear digital modulation with a known pulse shape or the OFDM modulation scheme. The linear modulation scheme is used in technologies such as CDMA, Wide-band CDMA (WCDMA) and Digital Video Broadcasting-Satellite (DVB-S) while technologies such as DVB-T, LTE, and WiMAX incorporate the OFDM modulation scheme.

Under the stated assumptions, we derive explicit closed-form expressions for a useful relaxed version of the GLRT for target detection in PMR networks depending on whether the noise variance is known or unknown. Numerical results show that the derived GLRTs perform significantly better than GLRTs that do not use the signal format information. Further, we observed the performance increases with the number of samples per symbol, and for a sufficiently large number of samples per symbol, the performance closely approximates that of an active radar where the transmitted signal is entirely known. Finally, the relaxation causes little loss at reasonable signal-to-noise ratios.

\textbf{Notations:} We use bold upper case, bold lower case, and italic lettering to respectively denote matrices, column vectors and scalars. Notations $(.)^T$, $(.)^H$ and $\otimes$ are the transpose, Hermitian and Kronecker product respectively. $\bm{I}_N$ stands for a $N$-dimensional identity matrix, $\bm{0}_{N\times1}$ denotes a column vector of length $N$ with all the elements equal to $0$ and $||.||$ is the Frobenius norm.

\section{Signal Model and Problem Statement}\label{section2}
We adopt the accepted model for PMR networks presented in \cite{Hack_PMR}. We assume $N_t$ transmit stations, $N_r$ receive stations and orthogonal (or separable) signals sent from each transmit station. The observations received directly from the transmitters are called reference channel signals, while those received from the possible reflection of the target are called surveillance channel signals. The reference and surveillance channel signals are separated using beamforming. After isolating the signals, each channel contains a certain amount of noise/clutter in addition to a scaled, delayed, and Doppler-shifted version of the transmitted signal. As in \cite{Hack_PMR}, we assume the delay-Doppler compensation accounts for the time delay and frequency shifts on the originally transmitted signal since we are testing for a target with a known position and Doppler. As in [1], we assume the noise/clutter has been whitened.

Let $\bm{s}_s^{ij} \in \mathbb{C}^{N \times 1}$ and $\bm{s}_r^{ij} \in \mathbb{C}^{N \times 1}$ denote the surveillance and reference channel signals, respectively, between the $i^{th}$ transmit station and $j^{th}$ receive station. These signals can be represented as
\begin{eqnarray}\label{basic_signal}
\bm{s}_s^{ij} & = \mu_s^{ij}\bm{u}^i + \bm{n}_s^{ij} \nonumber \\
\bm{s}_r^{ij} & = \mu_r^{ij}\bm{u}^i + \bm{n}_r^{ij},
\end{eqnarray}
where $\mu_{ij}^s$ and $\mu_{ij}^r$ are the unknown surveillance and reference channel coefficients, respectively, that include any gain due to beamforming and the noise vectors $\bm{n}^{r}_{ij}$ and $\bm{n}^{s}_{ij}$ are circular Gaussian noise, distributed as $\mathcal{CN}(\bm{0}_{N \times 1},\sigma^2\bm{I}_N)$ with $\sigma^2$ denoting the noise variance. Further, $\bm{u}_i \in \mathbb{C}^{N \times 1}$ contains samples of the unknown transmitted signal from the $i^{th}$ transmit station.

The PMR detection problem involves discriminating between the presence or absence of a target within a hypothesized Cartesian position-velocity cell under test \cite{Hack_PMR}. The problem can be formulated as a binary hypothesis test between the target-absent hypothesis ($\mathcal{H}_0$), and the target-present hypothesis ($\mathcal{H}_1$) as
\begin{eqnarray}\label{Hypotheses_test_PMR}
\mathcal{H}_0  : & \bm{s}_s^{ij} & = \bm{n}_s^{ij} \nonumber \\
& \bm{s}_r^{ij} & = \mu_r^{ij}\bm{u}^i + \bm{n}_r^{ij} \nonumber \\
\mathcal{H}_1 : & \bm{s}_s^{ij} & = \mu_s^{ij}\bm{u}^i + \bm{n}_s^{ij} \nonumber \\
 & \bm{s}_r^{ij} & = \mu^{ij}_r\bm{u}^i + \bm{n}_r^{ij},
\end{eqnarray}
for $i = 1, 2, \cdots, N_t$ and $j = 1, 2, \cdots, N_r$. In this paper, we only consider scenarios in which the transmitted signal vector $\bm{u}_i$, can be expressed as
\begin{eqnarray}\label{LinearModSignal}
\bm{u}^i & = & \bm{G}^i\bm{b}^i.
\end{eqnarray}
In (\ref{LinearModSignal}), $\bm{G}^i$ is a known matrix of appropriate size and $\bm{b}^i$ is a column vector of appropriate size containing unknown complex symbols from a digital modulation scheme. In the following subsections, we present communication signals that can be expressed in the form of (\ref{LinearModSignal}) along with the considered problem statement.

\subsection{Linear Digital Modulations}
\noindent
The complex baseband structure of a linear digital modulation scheme can be represented as \cite{proakis2008digital}
\begin{eqnarray}\label{LinearMod1}
u^i(t + nT_{sym}) & = & \sum_{k=0}^{M_i-1} g^i(t + kT_{sym}) b_{n - k}^i
\end{eqnarray}
for $0 \le t < T_{sym}$. In (\ref{LinearMod1}), $i$ denotes the index of the transmit station, $n$ denotes the symbol number index, $b_{k}^i$ denotes the transmitted complex baseband symbol, $T_{sym}$ is the symbol period of the digital modulation scheme and $g^i(.)$ denotes a pulse function of duration $M_iT_{sym}$ used at the $i^{th}$ transmit station. Popular pulse functions include the raised cosine and the root-raised cosine pulse shape \cite{Stinco}. After sampling, (\ref{LinearMod1}) can be rewritten as
\begin{eqnarray}\label{LinearMod2}
u^i(pT_s + nT_{sym}) & = & \sum_{k=0}^{M_i-1} g^i(pT_s + kT_{sym})b_{n-k}^i
\end{eqnarray}
for $p = 0, 1, \cdots, P-1$, where $P$ denotes the number of samples per symbol. In (\ref{LinearMod2}), $T_s = T_{sym}/P$ denotes the sampling interval. Collecting $N = LP$ samples from $L$ consecutive symbols indexed by $(n-L+1), (n-L+2), \cdots, n$, the transmitted signal samples can be expressed as
\begin{eqnarray}\label{LinearRep}
\bm{u}^i & = & \bm{G}^i\bm{b}^i,
\end{eqnarray}
where $\bm{u}^i = [(\bm{u}^i_{n})^T, (\bm{u}^i_{n-1})^T, \cdots, (\bm{u}^i_{n-L+1})^T]^T$ with $\bm{u}_k^i = [u^i(kT_{sym}), u^i(T_s + kT_{sym}), \cdots, u^i((P-1)T_s + kT_{sym})]^T$ for $k = (n-L+1), \cdots, n$ and $\bm{b}^{i} = \left[ b^i_{n}, b^i_{n-1}, \cdots, b^i_{n-L-M_i+2} \right]^T$. Let $\bm{G}^i$ be an $LP \times (L + M_i - 1)$ matrix defined as
\begin{eqnarray}\label{TransMatrix}
\bm{G}^i = 
\begin{bmatrix}
\bm{g}^i_{0} & \cdots & \bm{g}^i_{(M_i-1)} & \bm{0}_{P \times 1} & \cdots & \bm{0}_{P \times 1} \\
\bm{0}_{P \times 1} & \bm{g}^i_0  & \cdots & \bm{g}^i_{(M_i-1)} & \cdots & \bm{0}_{P \times 1} \\
\vdots & \ddots & \ddots & \ddots & \bm{0}_P & \vdots \\
\bm{0}_{P \times 1} & \bm{0}_{P \times 1} & \cdots & \bm{g}^i_0  & \cdots & \bm{g}^i_{(M_i-1)}
\end{bmatrix} 
\end{eqnarray}
where $\bm{g}^i_k = [ g^i(kT_{sym}), g^i(T_s + kT_{sym}), \cdots, g^i((P-1)T_s + kT_{sym})]^T$ for $k = 0, 1, \cdots, M_i-1$.

\subsection{Orthogonal Frequency-Division Multiplexing Signals}
\noindent
The complex baseband structure of an OFDM signal can be represented as \cite{Palmer}
\begin{eqnarray}\label{OFDM_sym}
u^i(t + nT_{sym}) & = & \sum_{l=0}^{N_s-1} e^{j2\pi \frac{l}{T_u}(t - T_g)}b^i_{nl},
\end{eqnarray}
for $0 \le t < T_{sym}$. In (\ref{OFDM_sym}), $i$ denotes the index of the transmit station, $n$ denotes the OFDM symbol number, $N_s$ is the number of subcarriers used in the OFDM signal, $b^i_{nl}$ are complex valued modulation symbols, $T_u$ is the duration of the useful part of the OFDM symbol (excluding the guard interval), $T_g$ is the guard interval duration, and $T_{sym} = (T_u + T_g)$ is the total OFDM symbol duration. Let $T_s$ be the sampling rate equal to $T_{sym}/(N_sP)$, where $P$ is the number of samples per complex symbol. Collecting $N = LN_sP$ samples from $L$ consecutive OFDM symbols indexed by $0, 1, \cdots, (L-1)$, the transmitted signal samples can be expressed as (similar to (\ref{LinearRep}))
\begin{eqnarray}\label{OFDMsym}
\bm{u}^i & = & (\bm{I}_L \otimes \bm{H})\bm{b}^i,
\end{eqnarray}
where $\bm{u}^i = [(\bm{u}^i_{0})^T, (\bm{u}^i_{1})^T, \cdots, (\bm{u}^i_{L - 1})^T]^T$ with $\bm{u}_k^i = [u^i(kT_{sym}), u^i(T_s + kT_{sym}), \cdots, u^i((N_sP-1)T_s + kT_{sym})]^T$ for $k = 0, 1, \cdots, L-1$ and $\bm{b}^i = [(\bm{b}_0^i)^T, (\bm{b}_1^i)^T, \cdots, (\bm{b}_{L-1}^i)^T]^T$ with $\bm{b}_k^i = [b_{k0}^i, b_{k1}^i, \cdots, b^i_{k(N_s-1)}  ]^T$ for $k = 0, 1, \cdots, L-1$. In (\ref{OFDMsym}), $\bm{H}$ is a $N_sP \times N_s$ matrix whose $ml^{th}$ element is given by
\begin{eqnarray}
h_{ml} & = & e^{\frac{j2\pi l(mT_s - T_g)}{T_u}}
\end{eqnarray} 
for $m = 0, 1, \cdots, N_sP-1$ and $l = 0, 1, \cdots, N_s-1$.


\subsection{Problem Statement}
\noindent
Under the stated assumptions, the PMR target detection problem in (\ref{Hypotheses_test_PMR}) can now be written as
\begin{eqnarray}\label{Hypotheses_test_linear}
\mathcal{H}_1: & \bm{s}_s^{ij} & = \mu^{ij}_s\bm{G}^i\bm{b}^i + \bm{n}_s^{ij}, \nonumber \\
& \bm{s}_r^{ij} & = \mu^{ij}_r\bm{G}^i\bm{b}^i + \bm{n}_r^{ij}, \nonumber \\
\mathcal{H}_0 : & \bm{s}_s^{ij} & = \bm{n}_s^{ij}, \nonumber \\
& \bm{s}_r^{ij} & = \mu^{ij}_r\bm{G}^i\bm{b}^i + \bm{n}_r^{ij},
\end{eqnarray}
for $i = 1, 2, \cdots, N_t$ and $j = 1, 2, \cdots, N_r$. In this paper, we derive a low complexity approximate GLRT for target detection in PMR networks that uses the available information regarding the signal format of the transmitted signal. We show significant detection performance improvement over the GLRT which ignores the signal format information.

\section{Target Detection in PMR networks}
Let $\bm{s}_{(s, r)}^{ij}$ denote the vector containing all the observations of all surveillance or reference signals associated with the $i^{th}$ transmit station and let $\bm{s}_{(s,r)}$ denote the concatenation of all $\bm{s}^i_{(s,r)}$, where the notation $(.)_{(s,r)}$ denotes either $(.)_s$ or $(.)_r$. We have
\begin{eqnarray}
\bm{s}_{(s,r)}^i & = & \left[ (\bm{s}_{(s,r)}^{i1})^T, \cdots,  (\bm{s}_{(s,r)}^{iN_r})^T\right]^T \in \mathbb{C}^{NN_r \times 1}, \nonumber \\
\bm{s}_{(s,r)} & = & \left[ (\bm{s}_{(s,r)}^1)^T, \cdots,  (\bm{s}_{(s,r)}^{N_t})^T\right]^T \in \mathbb{C}^{NN_rN_t \times 1}. \nonumber
\end{eqnarray}
Let $\bm{s}^i = \left[ (\bm{s}_s^i)^T, (\bm{s}_r^i)^T \right]^T$ and let $\bm{s} = [\bm{s}_s^T, \bm{s}_r^T]^T$ be the concatenation of all $\bm{s}^i$. Let $\bm{\mu}_{(s,r)}^i$ denote the vector of surveillance and reference channel coefficients associated with the $i^{th}$ transmit station and let $\bm{\mu}_{(s,r)}$ denote the concatenation of all $\bm{\mu}_{(s,r)}^i$ across the $N_t$ transmit stations defined as
\begin{eqnarray}
\bm{\mu}_{(s,r)}^i & = & \left[ {\mu}_{(s,r)}^{i1}, \cdots,  {\mu}_{(s,r)}^{iN_r}\right]^T \in \mathbb{C}^{N_r \times 1}, \nonumber \\
\bm{\mu}_{(s,r)} & = & \left[ (\bm{\mu}_{(s,r)}^1)^T, \cdots,  (\bm{\mu}_{(s,r)}^{N_t})^T\right]^T \in \mathbb{C}^{N_rN_t \times 1}. \nonumber
\end{eqnarray}
Finally, let $\bm{u} = \left[ (\bm{u}^1)^T, \cdots, (\bm{u}^{N_t})^T \right]^T \in \mathbb{C}^{N_tN \times 1}$ with $\bm{u}^i$ from (\ref{LinearModSignal}).

The received signals $\bm{s}_r^{ij}$ and $\bm{s}_s^{ij}$ in (\ref{basic_signal}) are parameterized by $\mu_r^{ij}$, $\mu_s^{ij}$ and $\bm{b}^i$. Since these parameters are unknown to the PMR system, we employ the GLRT for the hypotheses testing problem given in (\ref{Hypotheses_test_linear}). In GLRTs, we replace the unknown deterministic quantities with the corresponding maximum likelihood estimates (MLE). However, obtaining the MLE of the constellation symbols $b_k^i$ might not be tractable as we would have to search across all possible sequences of $\bm{b}^i$. Hence, we introduce a relaxation, called the relaxed GLRT, where we allow $b_k^i$ to be any complex number, i.e., $b_k^i \in \mathbb{C}$ as opposed to an actual modulation symbol from the defined finite set. Under this assumption, let $\bm{b}^i \in \mathbb{C}^{\mathcal{B}_i \times 1}$ and $\bm{b} = \left[ (\bm{b}^1)^T, \cdots, (\bm{b}^{N_t})^T \right]^T \in \mathbb{C}^{\mathcal{B} \times 1}$ with $\mathcal{B} = \sum_{i=1}^{N_t} \mathcal{B}_i$.

We now present a useful result along with some definitions that will be used extensively in the paper. Let $\bm{A}, \bm{B} \in \mathbb{C}^{K \times K}$ be Hermitian matrices with $\bm{A}$ being positive semidefinite and $\bm{B}$  positive definite. 
\begin{defn}
The generalized Hermitian eigenvalue problem is to compute a nonzero vector $\bm{w} \in \mathbb{C}^K$ and a real number $\lambda$ such that
\begin{eqnarray}
\bm{Aw} & = & \lambda \bm{Bw}
\end{eqnarray}
where $\bm{w}$ and the corresponding $\lambda$ are called the generalized eigenvector and generalized eigenvalue, respectively \cite{GHEP}.
\end{defn}
\begin{defn}
	The generalized Rayleigh quotient of the complex matrices $\bm{A}$ and $\bm{B}$ is a function of $\bm{w}$ and is defined as
	\begin{eqnarray}\label{Generalized_RayleighQuotient}
	R(\bm{w}) & = & \frac{\bm{w}^H\bm{Aw}}{\bm{w}^H\bm{Bw}}.
	\end{eqnarray}
\end{defn}

When $\bm{A}$, $\bm{B}$ are Hermitian matrices with $\bm{A}$ a positive semidefinite matrix and $\bm{B}$ a positive definite matrix, $R(\bm{w})$ has a maximum value equal to the largest generalized eigenvalue of $\bm{A}$ and $\bm{B}$ and the value of $\bm{w}$ that maximizes $R(\bm{w})$ is the generalized eigenvector of $\bm{A}$ and $\bm{B}$ corresponding to the largest generalized eigenvalue (See Section 4.4.3 of \cite{Alexander} or Theorem $5.24$ in \cite{Alexander}). In this paper, we denote $\lambda_1(\bm{A}, \bm{B})$ as the largest generalized eigenvalue of $\bm{A}$ and $\bm{B}$, and $v_1(\bm{A}, \bm{B})$ as the corresponding generalized eigenvector.

\subsection{Relaxed GLRT for PMR Networks When the Signal Format Information is Employed and $\sigma^2$ is Known}\label{sec3_ssec1}
\noindent
We consider the hypotheses testing problem given in (\ref{Hypotheses_test_linear}). The conditional probability density function (pdf) of $\bm{s}$ under $\mathcal{H}_1$ is given by
\begin{eqnarray}
p_1(\bm{s}|\bm{\mu}_s, \bm{\mu}_r, \bm{b}) & = & \prod_{i=1}^{N_t} p_1^i (\bm{s}^i| \bm{\mu}_s^i, \bm{\mu}_r^i, \bm{b}^i),
\end{eqnarray}
where
\begin{eqnarray}
p_1^i (\bm{s}^i| \bm{\mu}_s^i, \bm{\mu}_r^i, \bm{b}^i)  & \propto & \exp\bigg\{\frac{-1}{\sigma^2}\sum_{j=1}^{N_r} \bigg( ||\bm{s}_s^{ij} - \mu_s^{ij}\bm{G}^i\bm{b}^i||^2 \nonumber \\
&  & + ||\bm{s}_r^{ij} - \mu_r^{ij}\bm{G}^i\bm{b}^i||^2 \bigg) \bigg\}.
\end{eqnarray}
Similarly, the conditional pdf of $\bm{s}$ under $\mathcal{H}_0$ is given by
\begin{eqnarray}
p_0(\bm{s}|\bm{\mu}_r, \bm{b}) & = & \prod_{i=1}^{N_t} p_0^i (\bm{s}^i| \bm{\mu}_r^i, \bm{b}^i),
\end{eqnarray}
where
\begin{eqnarray}
p_0^i (\bm{s}^i| \bm{\mu}_r^i, \bm{b}^i) \propto  \exp\bigg\{\frac{-1}{\sigma^2}\sum_{j=1}^{N_r} ||\bm{s}_r^{ij} - \mu_r^{ij}\bm{G}^i\bm{b}^i||^2  \bigg\}.
\end{eqnarray}

Let $l_1(\bm{\mu}_s, \bm{\mu}_r, \bm{b}|\bm{s}) = \log p_1(\bm{s}|\bm{\mu}_s, \bm{\mu}_r, \bm{b})$ and $l_0(\bm{\mu}_r, \bm{b}|\bm{s}) = \log p_0(\bm{s}| \bm{\mu}_r, \bm{b})$ denote the log-likelihood functions under the hypotheses $\mathcal{H}_1$ and $\mathcal{H}_0$. The relaxed GLRT can now be written as
\begin{eqnarray}\label{GLRT_PMR_R}
\max_{\{\bm{\mu}_s, \bm{\mu}_r, \bm{b}\} \in \mathbb{C}^{N_rN_t} \times \mathbb{C}^{N_rN_t} \times \mathbb{C}^{\mathcal{B}} } l_1(\bm{\mu}_s, \bm{\mu}_r, \bm{b}|\bm{s}) \nonumber \\
- \max_{\{\bm{\mu}_r, \bm{b}\} \in \mathbb{C}^{N_rN_t} \times \mathbb{C}^{\mathcal{B}} } l_0( \bm{\mu}_r, \bm{b}|\bm{s}) \LRT{\mathcal{H}_1}{\mathcal{H}_0} \kappa_{ksf},
\end{eqnarray}
where $\kappa_{ksf}$ denotes a threshold corresponding to a desired value of false alarm probability. It is shown in Appendix \ref{pmr_glrt_with_sig_noise_info} that the GLRT-based target detector in (\ref{GLRT_PMR_R}), termed the \emph{Passive MIMO Radar Relaxed GLRT with Known signal format and known noise variance (PMR-RGLRT-K)}, is given by 
\begin{eqnarray}
\xi_{ksf} & = &\frac{1}{\sigma^2}\sum_{i=1}^{N_t} \Big[ \lambda_1\left((\bm{G}^i)^H\bm{\phi}_1^i (\bm{\phi}_1^i)^H\bm{G}^i,  (\bm{G}^i)^H \bm{G}^i \right) \nonumber \\
&&- \lambda_1\left((\bm{G}^i)^H\bm{\phi}_r^i (\bm{\phi}_r^i)^H \bm{G}^i,  (\bm{G}^i)^H \bm{G}^i \right) \Big] \nonumber \\
&&\LRT{\mathcal{H}_1}{\mathcal{H}_0} \kappa_{ksf} 
\end{eqnarray}
where $\bm{\phi}_1^i = [\bm{\phi}_s^i, \bm{\phi}_r^i]$, and the matrices $\bm{\phi}_s^i$ and $\bm{\phi}_r^i$ are defined as
\begin{eqnarray}
\bm{\phi}^i_{(s, r)} & = & \left[ \bm{s}_{(s, r)}^{i1}, \bm{s}_{(s, r)}^{i2}, \cdots, \bm{s}_{(s, r)}^{iN_r} \right] \in \mathbb{C}^{N \times N_r}.
\end{eqnarray}

In specific scenarios discussed in \cite{Hack_noRef, Liu}, the direct path reference channel signals might not be available in the PMR networks. The target detection problem in such scenarios, termed as Passive Source Localization (PSL) networks, can be formulated as 
\begin{eqnarray}\label{Hypotheses_test_PSL}
\mathcal{H}_0  : & \bm{s}_s^{ij} & = \bm{n}_s^{ij} \nonumber \\
\mathcal{H}_1 : & \bm{s}_s^{ij} & = \mu^{ij}_s\bm{G}^i\bm{b}^i + \bm{n}_s^{ij},
\end{eqnarray}
for $i = 1, 2, \cdots, N_t$ and $j = 1, 2, \cdots, N_r$. The PSL and PMR hypotheses tests are equivalent if the PMR system ignores the direct-path reference channel signals $\bm{s}_r^{ij}$ \cite{Hack_PMR}. It is shown in Appendix \ref{psl_glrt_sig_noise_info}  that the relaxed GLRT-based target detector that uses the signal structure information, termed the \emph{Passive Source Localization Relaxed GLRT with Known signal format and known noise variance (PSL-RGLRT-K)}, for the hypotheses testing problem in (\ref{Hypotheses_test_PSL}), is given by 
\begin{eqnarray}
\xi_{psk} = \frac{1}{\sigma^2}\sum_{i=1}^{N_t} \lambda_1\left((\bm{G}^i)^H \bm{\phi}_s^i (\bm{\phi}_s^i)^H \bm{G}^i,  (\bm{G}^i)^H \bm{G}^i \right) \LRT{\mathcal{H}_1}{\mathcal{H}_0} \kappa_{psk}, 
\end{eqnarray}
where $\kappa_{psk}$ denotes a threshold corresponding to a desired value of false alarm probability.

\subsection{GLRT When the Signal Format Information is Employed and $\sigma^2$ is Unknown}\label{sec3_ssec2}
\noindent
When $\sigma^2$ is unknown, the conditional pdf of $\bm{s}$ under $\mathcal{H}_1$ is given by
\begin{eqnarray}
p_1(\bm{s}|\bm{\mu}_s, \bm{\mu}_r, \bm{b}, \sigma^2) & = & \prod_{i=1}^{N_t} p_1^i (\bm{s}^i| \bm{\mu}_s^i, \bm{\mu}_r^i, \bm{b}^i, \sigma^2)
\end{eqnarray}
where, 
\begin{flalign*}
p_1^i (\bm{s}^i| \bm{\mu}_s^i, \bm{\mu}_r^i, \bm{b}^i, \sigma^2) &  &
\end{flalign*}
\begin{eqnarray}
= &  \frac{1}{(\pi\sigma^2)^{N_rN}} \exp\bigg\{\frac{-1}{\sigma^2}  \sum_{j=1}^{N_r} \big(  ||\bm{s}_s^{ij} - \mu_s^{ij}\bm{G}^i\bm{b}^i||^2 \nonumber \\
&  + ||\bm{s}_r^{ij} - \mu_r^{ij}\bm{G}^i\bm{b}^i||^2\big) \bigg\}.
\end{eqnarray}
The conditional pdf of $\bm{s}$ under $\mathcal{H}_0$, $p_0(\bm{s}|\bm{\mu}_s, \bm{\mu}_r, \bm{b}, \sigma^2)$, is similarly defined. Let $l_1(\bm{\mu}_s, \bm{\mu}_r, \bm{b}, \sigma^2|\bm{s}) = \log p_1(\bm{s}|\bm{\mu}_s, \bm{\mu}_r, \bm{b}, \sigma^2)$ and $l_0(\bm{\mu}_r, \bm{b}, \sigma^2|\bm{s}) = \log p_0(\bm{s}| \bm{\mu}_r, \bm{b}, \sigma^2)$ denote the log-likelihood functions under the hypotheses $\mathcal{H}_1$ and $\mathcal{H}_0$. The relaxed GLRT is given by
\begin{eqnarray}\label{GLRT_PMR_UK}
\max_{\{\bm{\mu}_s, \bm{\mu}_r, \bm{b}, \sigma^2\} \in \mathbb{C}^{N_rN_t} \times \mathbb{C}^{N_rN_t} \times \mathbb{C}^{\mathcal{B}} \times \mathbb{R}^+} l_1(\bm{\mu}_s, \bm{\mu}_r, \bm{b}, \sigma^2|\bm{s}) \nonumber \\
- \max_{\{\bm{\mu}_r, \bm{b}, \sigma^2\} \in \mathbb{C}^{N_rN_t} \times \mathbb{C}^{\mathcal{B}} \times \mathbb{R}^+} l_0( \bm{\mu}_r, \bm{b}, \sigma^2|\bm{s}) \LRT{\mathcal{H}_1}{\mathcal{H}_0} \kappa_{uk},
\end{eqnarray}
where $\kappa_{uk}$ denotes a threshold corresponding to a desired value of false alarm probability. It is shown in Appendix \ref{pmr_glrt_with_sig_info} that the GLRT-based target detector in (\ref{GLRT_PMR_UK}),  termed the \emph{Passive MIMO Radar Relaxed GLRT with unknown noise variance and Known signal format  (PMR-RGLRT-UK)}, is given by 
\begin{eqnarray}
\xi_{uk} & = & \frac{\sum_{i=1}^{N_t} \Big[ E_{sr}^i - \lambda_1\left((\bm{G}^i)^H\bm{\phi}_r^i (\bm{\phi}_r^i)^H\bm{G}^i,  (\bm{G}^i)^H\bm{G}^i \right) \Big]}{\sum_{i=1}^{N_t} \Big[ E_{sr}^i - \lambda_1\left((\bm{G}^i)^H\bm{\phi}_1^i (\bm{\phi}_1^i)^H\bm{G}^i,  (\bm{G}^i)^H\bm{G}^i \right) \Big]} \nonumber \\
& & \LRT{\mathcal{H}_1}{\mathcal{H}_0} \kappa_{uk},
\end{eqnarray}
where the scalar $E_{sr}^i = ||\bm{s}_s^i||^2 + ||\bm{s}_r^i||^2$.

\section{Simulation Results}
In this section, we compare the performance of the proposed GLRT-based target detectors to other GLRT-based  detectors available in the literature via numerical simulations. We briefly describe the considered GLRT-based target detectors:
\subsubsection{Active (known signal) MIMO Radar GLRT (AMR-GLRT)}
The binary hypothesis test between the target-absent hypothesis ($\mathcal{H}_0$), and the target-present hypothesis ($\mathcal{H}_1$) in an active radar network (where the transmitted signals are known) can be formulated as
\begin{eqnarray}\label{Hypotheses_test_AMR}
\mathcal{H}_0  : & \bm{s}_s^{ij} & = \bm{n}_s^{ij} \nonumber \\
\mathcal{H}_1 : & \bm{s}_s^{ij} & = \mu_s^{ij}\bm{u}^i + \bm{n}_s^{ij},
\end{eqnarray}
for $i = 1, 2, \cdots, N_t$ and $j = 1, 2, \cdots, N_r$, where the transmitted signal $\bm{u}^i$ is assumed known and the channel coefficients ${\mu}_s^{ij}$ are deterministic unknowns. The GLRT for (\ref{Hypotheses_test_AMR}) is given by \cite{AMRGLRT}
\begin{eqnarray}
\xi_{amr} & = & \frac{1}{\sigma^2}\sum_{i=1}^{N_t}\sum_{j=1}^{N_r} |(\bm{u}^i)^H\bm{s}_s^{ij}|^2 \LRT{\mathcal{H}_1}{\mathcal{H}_0} \kappa_{amr},
\end{eqnarray}
where $\kappa_{amr}$ denotes a threshold corresponding to a desired false alarm probability. 

\subsubsection{Passive MIMO Radar GLRT without using the signal format information (PMR-GLRT)}
The GLRT for target detection in PMR networks which does not employ knowledge of the signal format for the hypotheses testing problem given in (\ref{Hypotheses_test_PMR}) was derived in \cite{Hack_PMR} and is given by
\begin{eqnarray}
\xi_{pmr} & = & \frac{1}{\sigma^2}\sum_{i=1}^{N_t}  \bigg[\lambda_1^*\left(\bm{\phi}_1^i (\bm{\phi}_1^i)^H\right) - \lambda_1^*\left(\bm{\phi}_r^i (\bm{\phi}_r^i)^H \right) \bigg] \nonumber \\
& & \LRT{\mathcal{H}_1}{\mathcal{H}_0} \kappa_{pmr}, 
\end{eqnarray}
where $\kappa_{pmr}$ denotes a threshold corresponding to a desired false alarm probability and $\lambda_1^*(\bm{A})$ denotes the largest eigenvalue of matrix $\bm{A}$. 

\subsubsection{Passive Source Localization GLRT without using the signal format information (PSL-GLRT)}
The GLRT for target detection in PSL networks which does not employ knowledge of the signal format for the hypotheses testing problem given in (\ref{Hypotheses_test_PSL}) was derived in \cite{Hack_noRef} and is given by
\begin{eqnarray}
\xi_{psl} & = & \frac{1}{\sigma^2}\sum_{i=1}^{N_t} \lambda_1^*\left(\bm{\phi}_s^i (\bm{\phi}_s^i)^H\right) \LRT{\mathcal{H}_1}{\mathcal{H}_0} \kappa_{psl},
\end{eqnarray}
where $\kappa_{psl}$ denotes a threshold corresponding to a desired false alarm probability. Table \ref{sec4_table1} provides the test statistics of the various considered GLRT-based detectors.
\begin{table*}[t]
	\begin{center}
		\begin{tabular}{|c|c|c|}
			\hline
			{\bf Abbreviation} & {\bf Test Statistic}  & {\bf Corresponding References} \\
			\hline 
			\emph{AMR-GLRT} & $\frac{1}{\sigma^2}\sum_{i=1}^{N_t}\sum_{j=1}^{N_r} |(\bm{u}^i)^H\bm{s}_s^{ij}|^2 $ & \cite{AMRGLRT} \\
			\hline
			\emph{PMR-GLRT} & $\frac{1}{\sigma^2}\sum_{i=1}^{N_t}  \bigg[\lambda_1^*\left(\bm{\phi}_1^i (\bm{\phi}_1^i)^H\right) - \lambda_1^*\left(\bm{\phi}_r^i (\bm{\phi}_r^i)^H \right) \bigg]$ & \cite{Hack_PMR, Hack_PMR_conf} \\
			\hline
			\emph{PSL-GLRT} & $\frac{1}{\sigma^2}\sum_{i=1}^{N_t} \lambda_1^*\left(\bm{\phi}_s^i (\bm{\phi}_s^i)^H\right)$ & \cite{Hack_noRef} \\
			\hline			
			\emph{PMR-RGLRT-K} & $\frac{1}{\sigma^2}\sum_{i=1}^{N_t}  \bigg[\lambda_1\left((\bm{G}^i)^H\bm{\phi}_1^i (\bm{\phi}_1^i)^H\bm{G}^i,  (\bm{G}^i)^H\bm{G}^i \right) 
			 - \lambda_1\left((\bm{G}^i)^H\bm{\phi}_r^i (\bm{\phi}_r^i)^H\bm{G}^i,  (\bm{G}^i)^H\bm{G}^i \right) \bigg]$ & Proposed in this paper \\
			\hline	
			\emph{PSL-RGLRT-K} & $\frac{1}{\sigma^2}\sum_{i=1}^{N_t}  \bigg[\lambda_1\left((\bm{G}^i)^H\bm{\phi}_s^i (\bm{\phi}_s^i)^H\bm{G}^i,  (\bm{G}^i)^H\bm{G}^i \right)$ & Proposed in this paper \\
			\hline				
			\emph{PMR-RGLRT-UK} & $ \frac{\sum_{i=1}^{N_t} \Big[ E^i - \lambda_1\left((\bm{G}^i)^H\bm{\phi}_r^i (\bm{\phi}_r^i)^H\bm{G}^i,  (\bm{G}^i)^H\bm{G}^i \right) \Big]}{\sum_{i=1}^{N_t} \Big[ E^i - \lambda_1\left((\bm{G}^i)^H\bm{\phi}_1^i (\bm{\phi}_1^i)^H\bm{G}^i,  (\bm{G}^i)^H\bm{G}^i \right) \Big]} $ & Proposed in this paper \\			
			\hline			
		\end{tabular}
		\caption{Test statistics of various GLRT target detectors.}\label{sec4_table1}
	\end{center}
\end{table*}

\subsection{Simulation scenario}\label{sim_setup}
\noindent
For a fair comparison, we follow the simulation setup of \cite{Hack_PMR}. We consider a PMR network with $N_t = 2$ transmit stations and $N_r = 3$ receive stations. Following \cite{Hack_PMR}, we fix $||\bm{u}^i||^2 = N$. The transmitted signal samples $\bm{u}^i$ are generated according to the chosen signal format in (\ref{LinearModSignal}) across all transmit stations. The reference and surveillance signal samples are generated on each Monte Carlo trial according to the signal model given in (\ref{basic_signal}). As in \cite{Hack_PMR}, the reference channel coefficients, $\bm{\mu}_r^i$, are randomly drawn from a $\mathcal{CN}(\bm{0}_{N_r}, \bm{I}_{N_r})$ distribution on each trial under $\mathcal{H}_0$ and $\mathcal{H}_1$, and then scaled to achieve a desired direct-path signal-to-noise ratio ($\mbox{DNR}^i_{{avg}}$) according to  
\begin{eqnarray}
\mbox{DNR}^i_{{avg}} & = & \frac{||\bm{\mu}_r^i||^2}{N_r\sigma^2}
\end{eqnarray}
on each trial, where $\bm{\mu}_r^i = [\mu_r^{i1}, \cdots, \mu_r^{iN_r}]^T$ and $|\mu_r^{ij}|^2/\sigma^2$ is the DNR of the $ij^{th}$ reference channel. Surveillance channel coefficients are similarly drawn from a $\mathcal{CN}(\bm{0}_{N_r}, \bm{I}_{N_r})$ distribution and scaled to achieve a desired surveillance signal-to-noise ratio ($\mbox{SNR}^i_{{avg}}$) according to  
\begin{eqnarray}
\mbox{SNR}^i_{{avg}} & = & \frac{||\bm{\mu}_s^i||^2}{N_r\sigma^2}
\end{eqnarray}
on each trial, where $\bm{\mu}_r^i = [\mu_r^{i1}, \cdots, \mu_r^{iN_r}]^T$ and $|\mu_s^{ij}|^2/\sigma^2$ is the SNR of the $ij^{th}$ surveillance channel. For simplicity, we assume that $\mbox{SNR}^i_{{avg}} = \mbox{SNR}_{{avg}}$ for all $i$, i.e., the average surveillance channel target-path SNR across receivers is the same for each transmit channel. Similarly, we assume $\mbox{DNR}^i_{{avg}} = \mbox{DNR}_{{avg}}$ and $\bm{G}^i(.) = \bm{G}(.)$ for all $i$. In our simulations, we consider cases where the transmitted signal is either linearly modulated or follows the OFDM modulation scheme.

The transmitted signal is generated according to (\ref{LinearMod1}) in case of the linear modulations. The complex baseband symbols are chosen from a Binary Phase Shift Keying (BPSK) constellation and $g^i(.)$ is a raised cosine pulse of roll-off factor $0.22$ and duration $8T_{sym}$. When the transmitted signal uses the OFDM modulation, it is generated according to (\ref{OFDM_sym}). The number of sub-carriers in the OFDM symbol is $16$, the guard-interval duration $T_g$ is $0$ $\mu s$ and BPSK symbols are modulated on each sub-carrier of the OFDM symbol.

For all the considered target detectors, the detection threshold that achieves a probability of false alarm $(P_f)$ of $10^{-3}$ is determined empirically using $10^4$ trials under $\mathcal{H}_0$, and the probability of detection ($P_d$) is estimated using $10^4$ trials under $\mathcal{H}_1$. The number of symbols used for target detection in the case of the linearly modulated transmitted signal is $10$ (total of $10P$ samples), while in the case of the OFDM modulated transmitted signal, we use $1$ OFDM symbol (total of $16P$ samples). The BPSK symbols used in the generation of the transmitted signal are randomly generated for each Monte-Carlo simulation run.

\subsection{Numerical results}
\noindent
\subsubsection{Dependence on $\mbox{SNR}_{avg}$, $\mbox{DNR}_{avg}$ and $P$ }
\begin{figure}[t]
	\centering
	\begin{subfigure}[b]{\columnwidth}
		\centering
		\includegraphics[height = 2.2 in, width = \columnwidth]{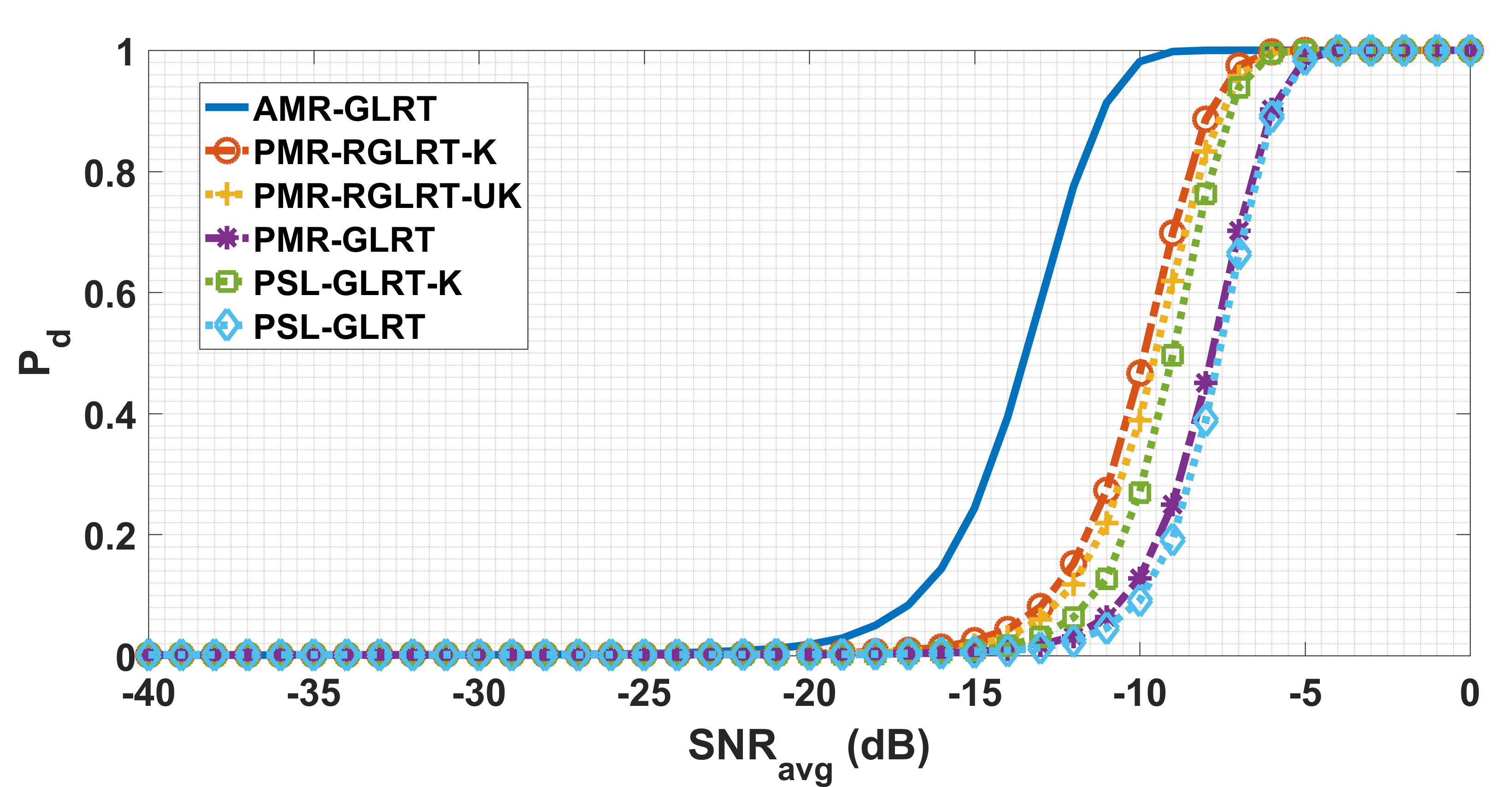}
		\caption{ }
	\end{subfigure}	
	~ 
	\begin{subfigure}[b]{\columnwidth}
		\centering
		\includegraphics[height = 2.2 in, width = \columnwidth]{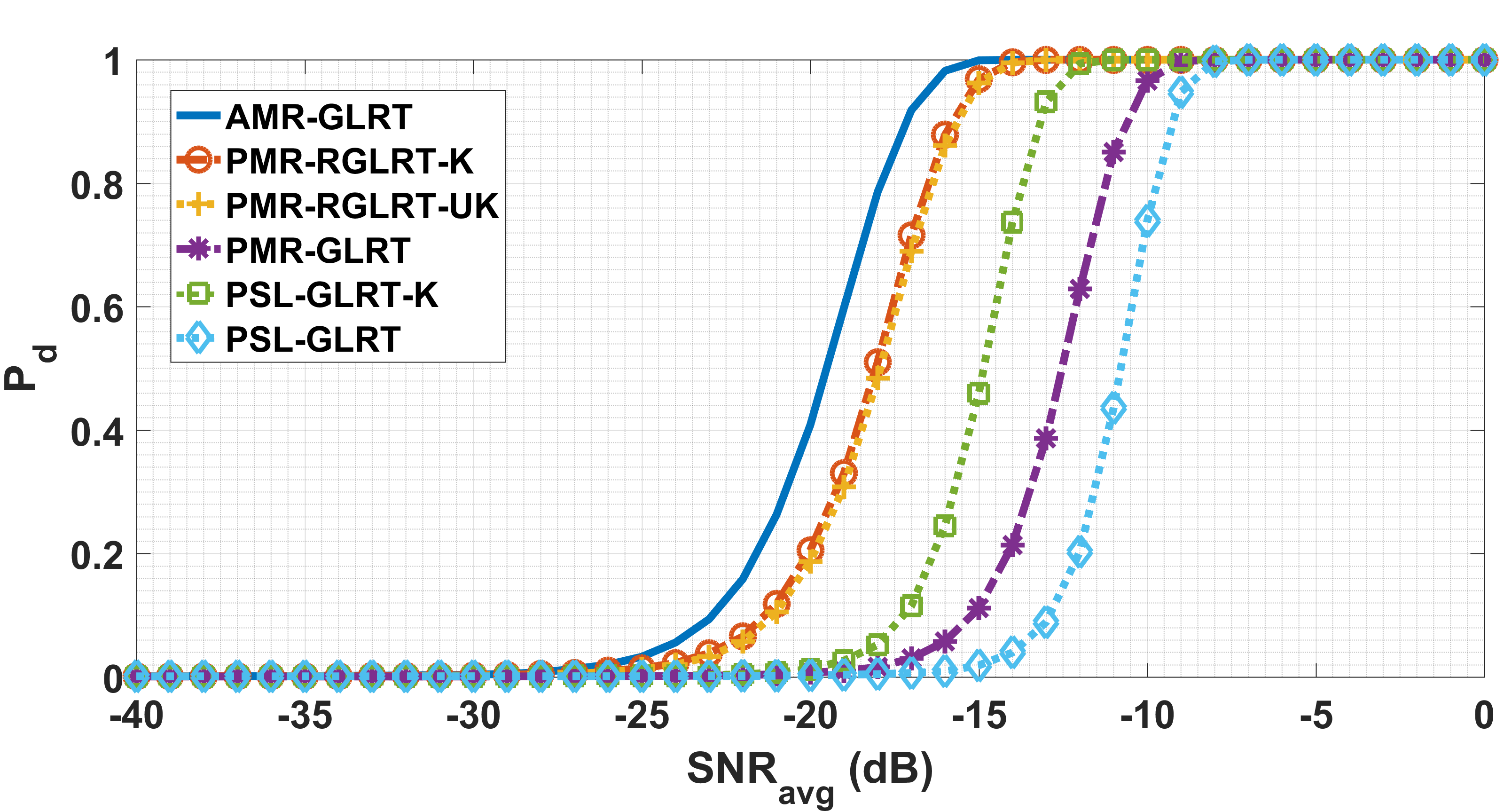}
		\caption{ }
	\end{subfigure}	
	
	~ 
	\begin{subfigure}[b]{\columnwidth}
		\centering
		\includegraphics[height = 2.2 in, width = \columnwidth]{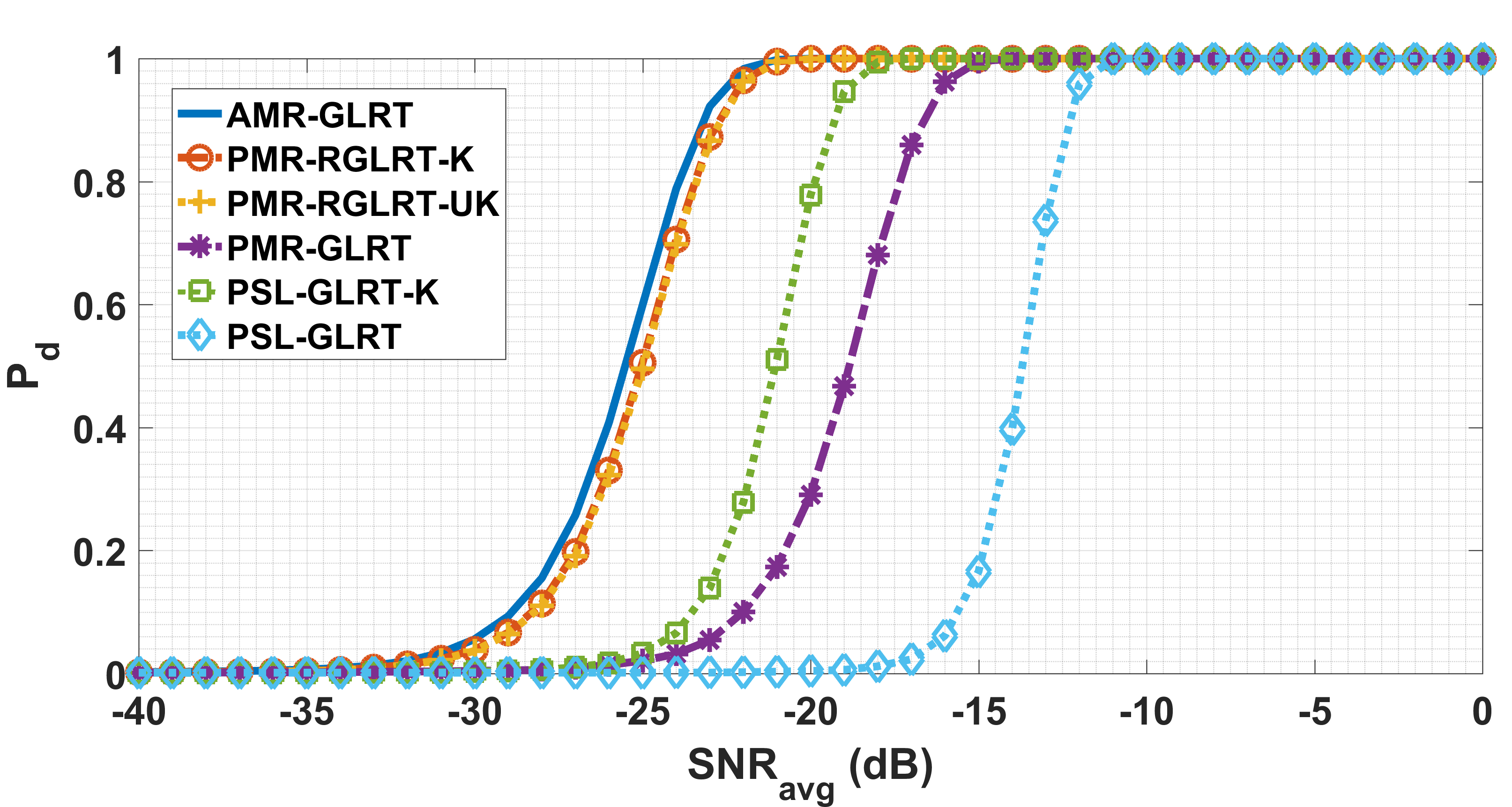}
		\caption{ }
	\end{subfigure}		
	
	\caption{$P_d$ curves as a function of $\mbox{SNR}_{avg}$ when the transmitted signal is linearly modulated with BPSK symbols, $L = 10$ symbols and $\mbox{DNR}_{avg} = -10$ dB for different values of samples per symbol $P$, (a) $P = 4$, (b) $P = 16$, (c) $P = 64$.}\label{Pd_bpsk_results_dnr_ne10dB}
\end{figure}

\begin{figure}[t]
	\centering
	\begin{subfigure}[b]{\columnwidth}
		\centering
		\includegraphics[height = 2.2 in, width = \columnwidth]{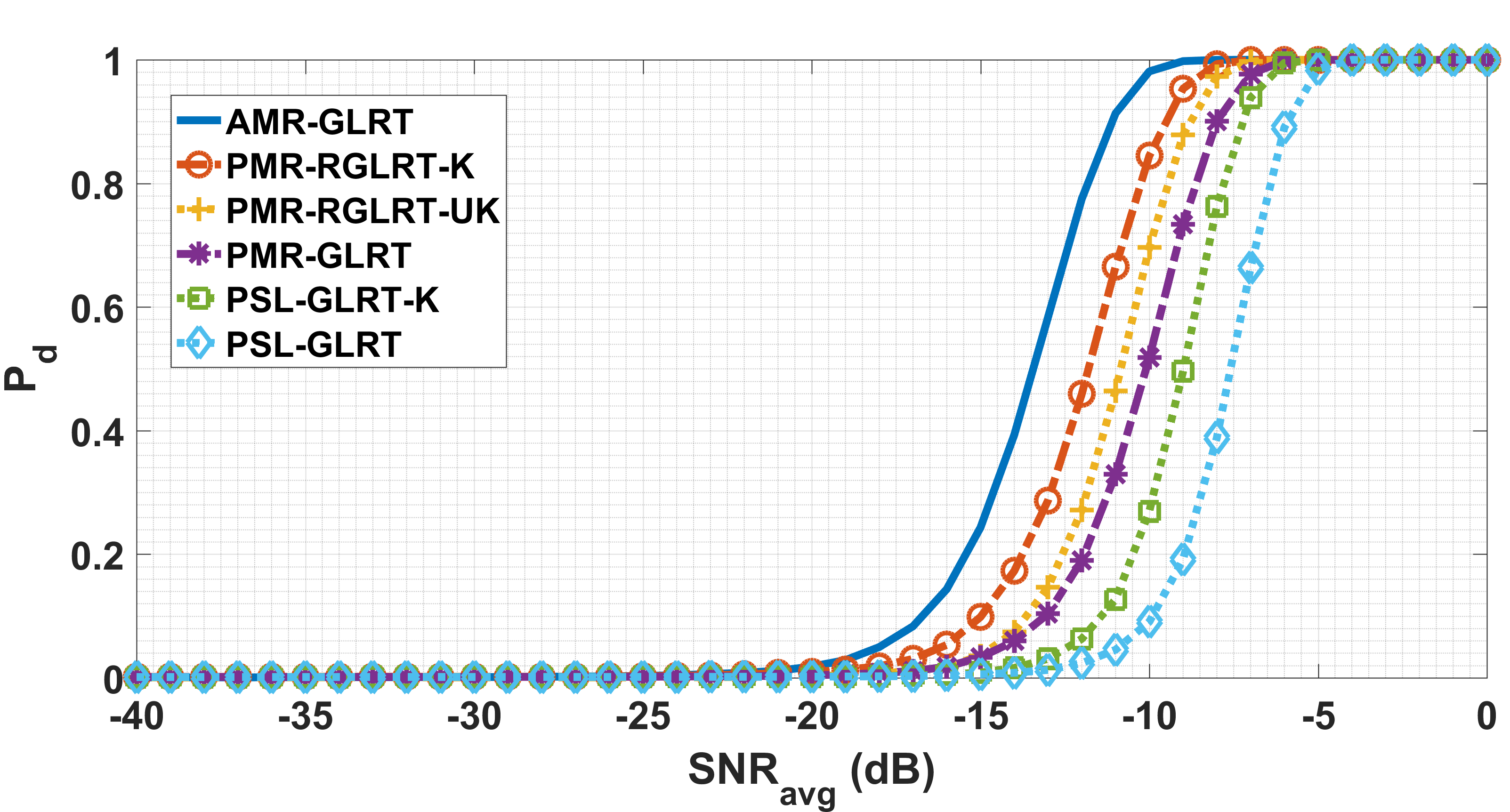}
		\caption{ }
	\end{subfigure}	
	~ 
	\begin{subfigure}[b]{\columnwidth}
		\centering
		\includegraphics[height = 2.2 in, width = \columnwidth]{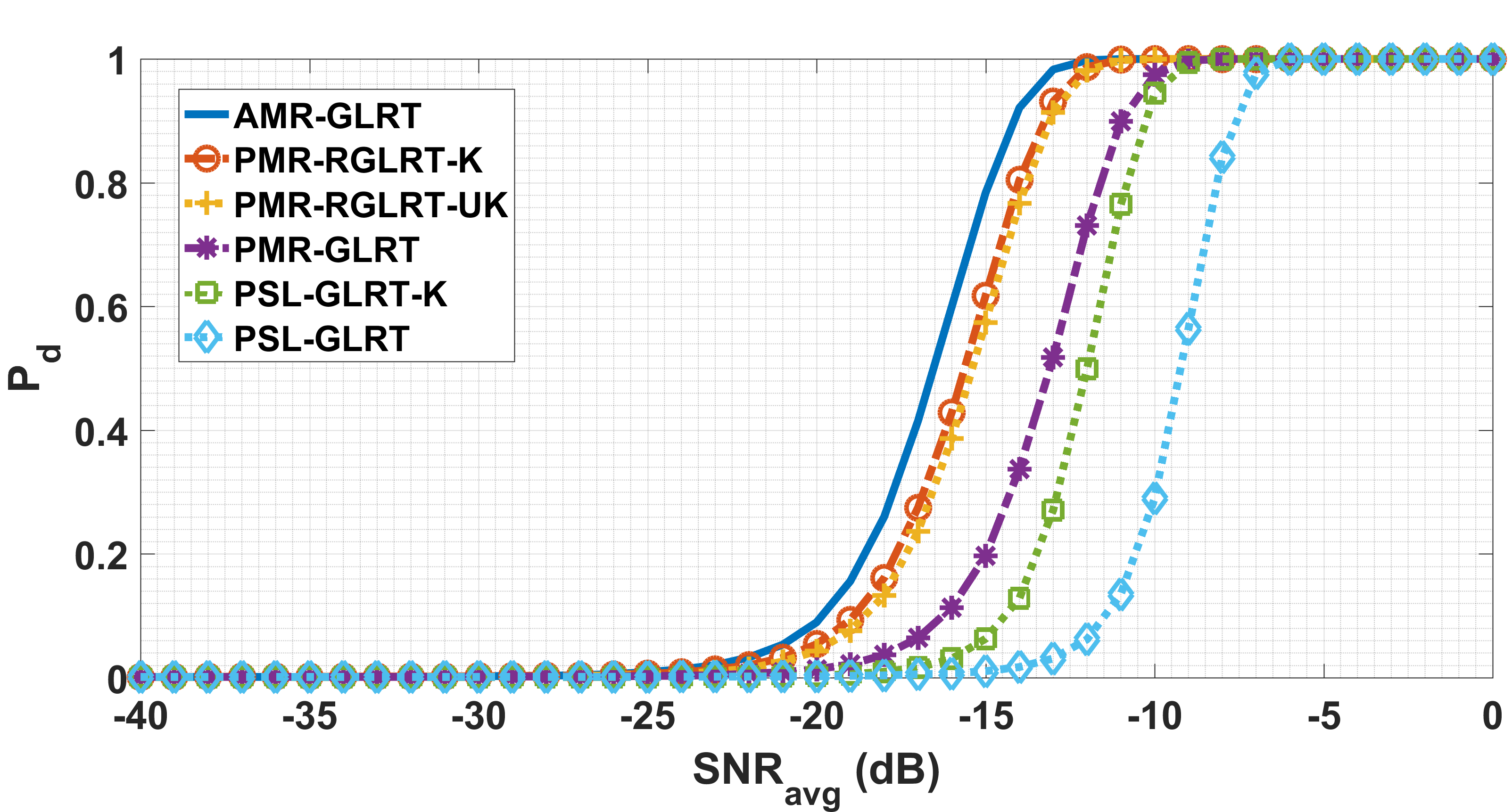}
		\caption{ }
	\end{subfigure}	
	
	~ 
	\begin{subfigure}[b]{\columnwidth}
		\centering
		\includegraphics[height = 2.2 in, width = \columnwidth]{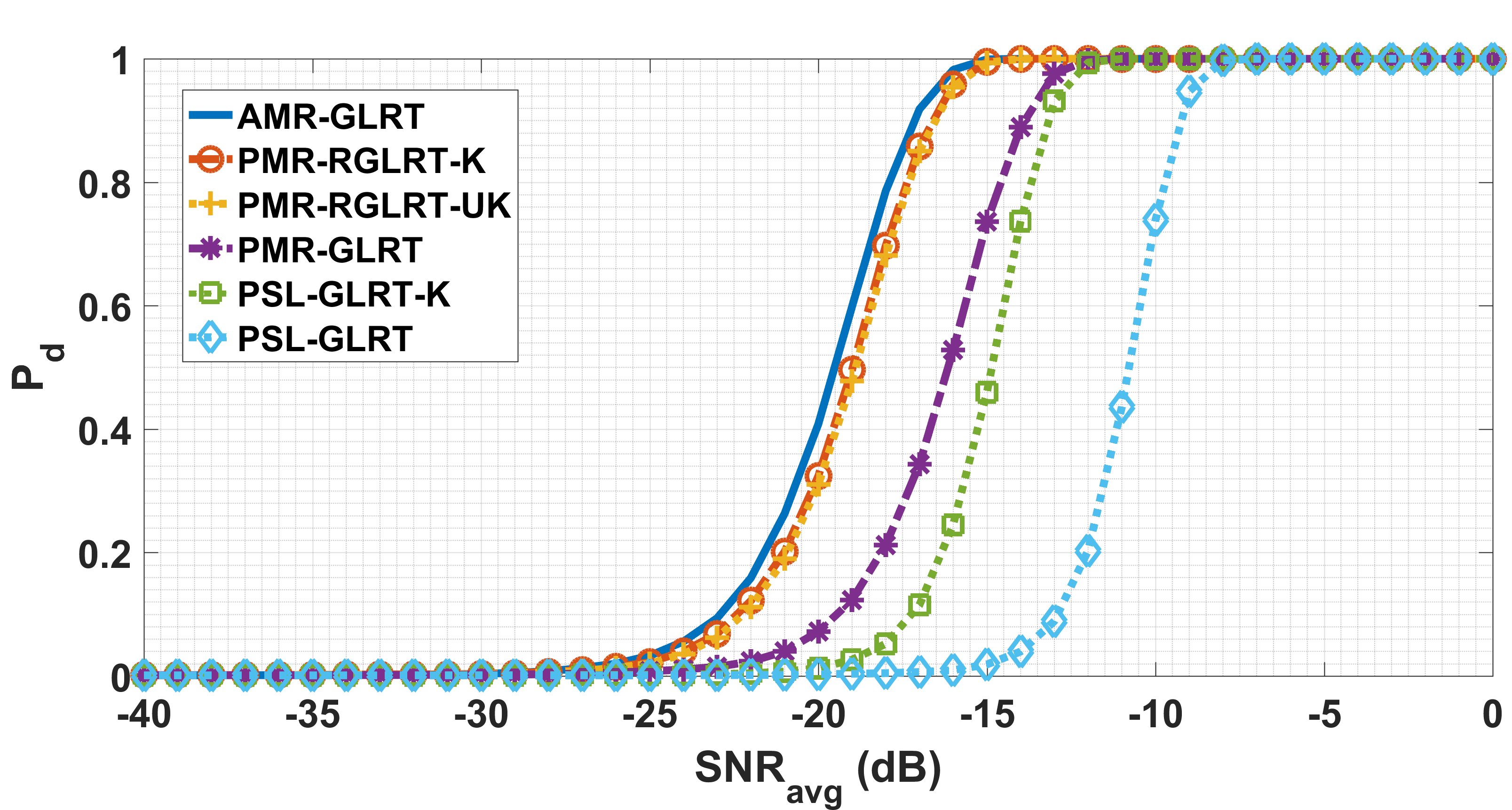}
		\caption{ }
	\end{subfigure}		
	
	\caption{$P_d$ curves as a function of $\mbox{SNR}_{avg}$ when the transmitted signal is linearly modulated with BPSK symbols, $L = 10$ symbols and $\mbox{DNR}_{avg} = -5$ dB for different values of samples per symbol $P$, (a) $P = 4$, (b) $P = 8$, (c) $P = 16$.}\label{Pd_bpsk_results_dnr_ne5dB}
\end{figure}

\begin{figure}[t]
	\centering
	\begin{subfigure}[b]{\columnwidth}
		\centering
		\includegraphics[height = 2.2 in, width = \columnwidth]{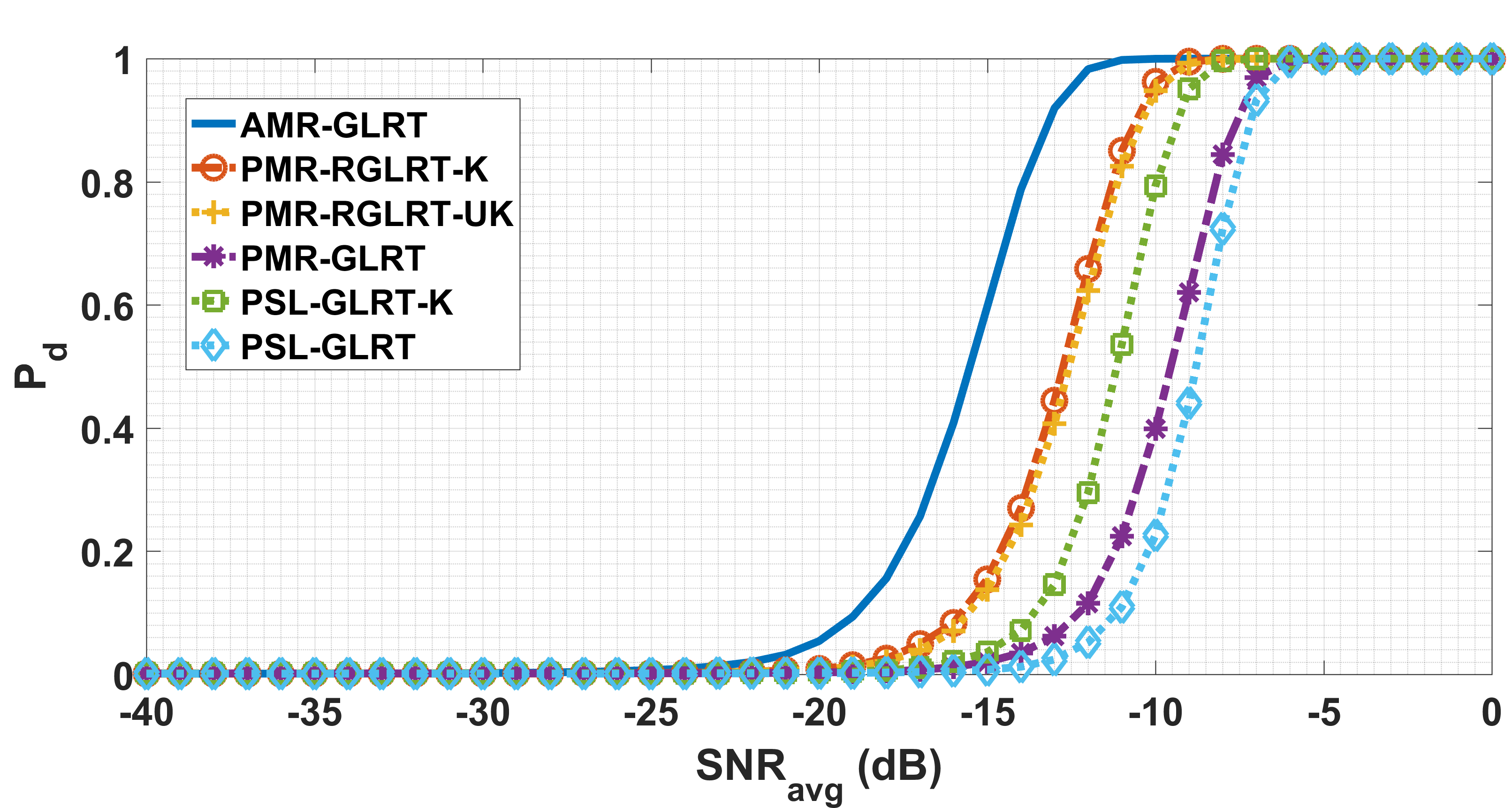}
		\caption{ }
	\end{subfigure}	
	~ 
	\begin{subfigure}[b]{\columnwidth}
		\centering
		\includegraphics[height = 2.2 in, width = \columnwidth]{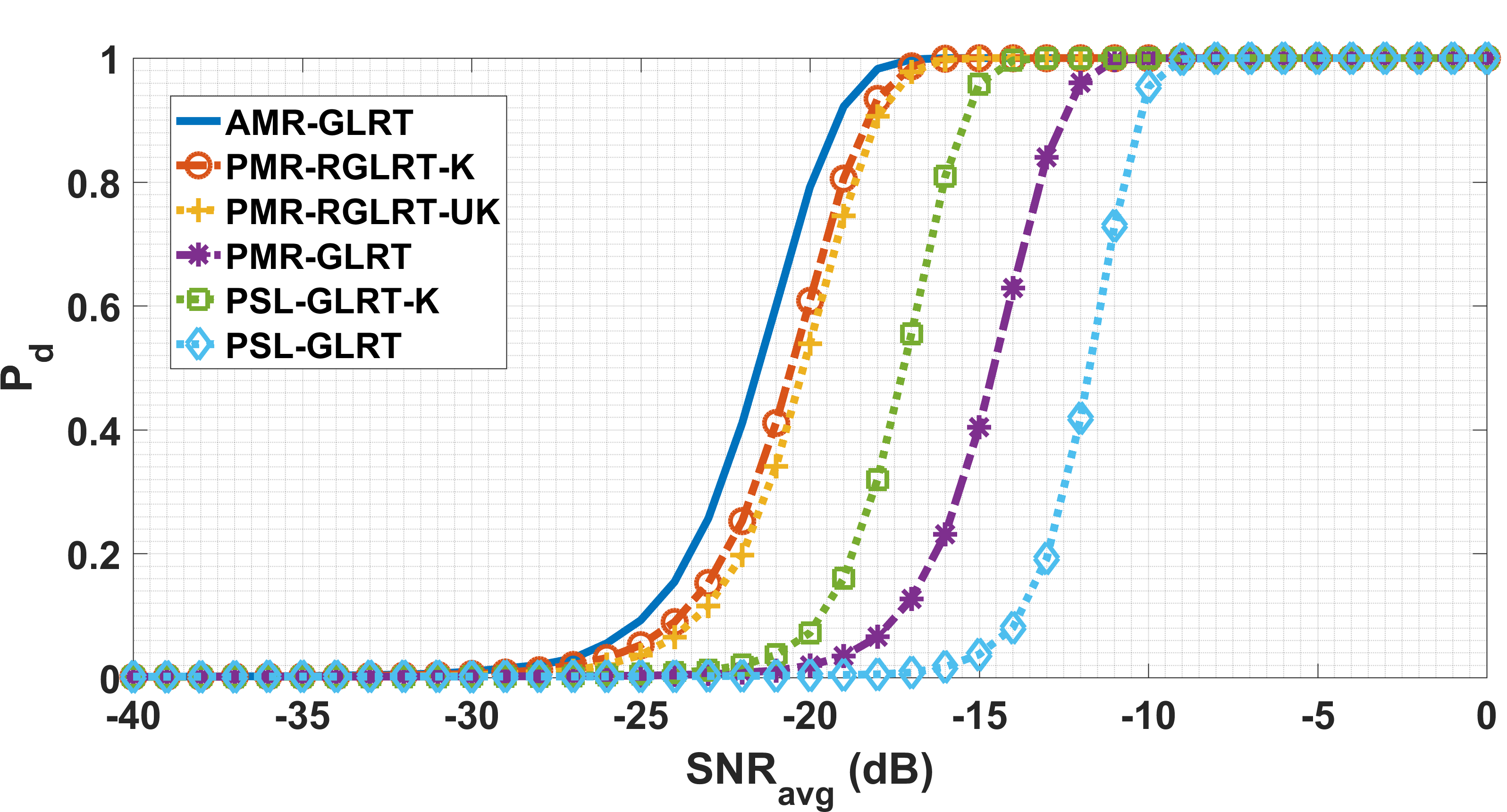}
		\caption{ }
	\end{subfigure}	
	
	~ 
	\begin{subfigure}[b]{\columnwidth}
		\centering
		\includegraphics[height = 2.2 in, width = \columnwidth]{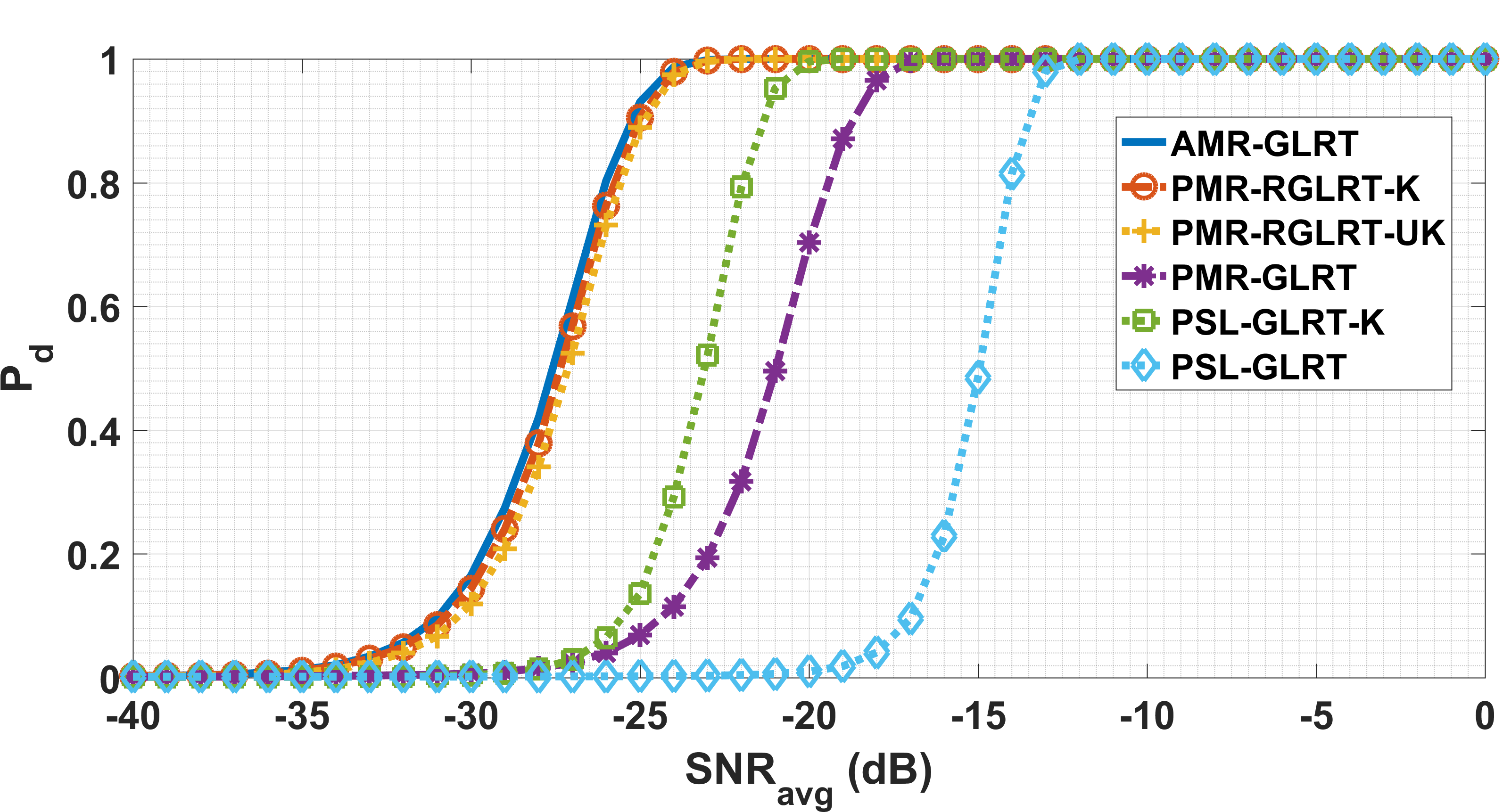}
		\caption{ }
	\end{subfigure}		
	
	\caption{$P_d$ curves as a function of $\mbox{SNR}_{avg}$ when the transmitted signal is an OFDM signal with $N_s = 16$ subcarriers and $\mbox{DNR}_{avg} = -10$ dB for different values of samples per symbol $P$, (a) $P = 4$, (b) $P = 16$, (c) $P = 64$.}\label{Pd_ofdm_results_dnr_ne10dB}
\end{figure}

\begin{figure}[t]
	\centering
	\begin{subfigure}[b]{\columnwidth}
		\centering
		\includegraphics[height = 2.2 in, width = \columnwidth]{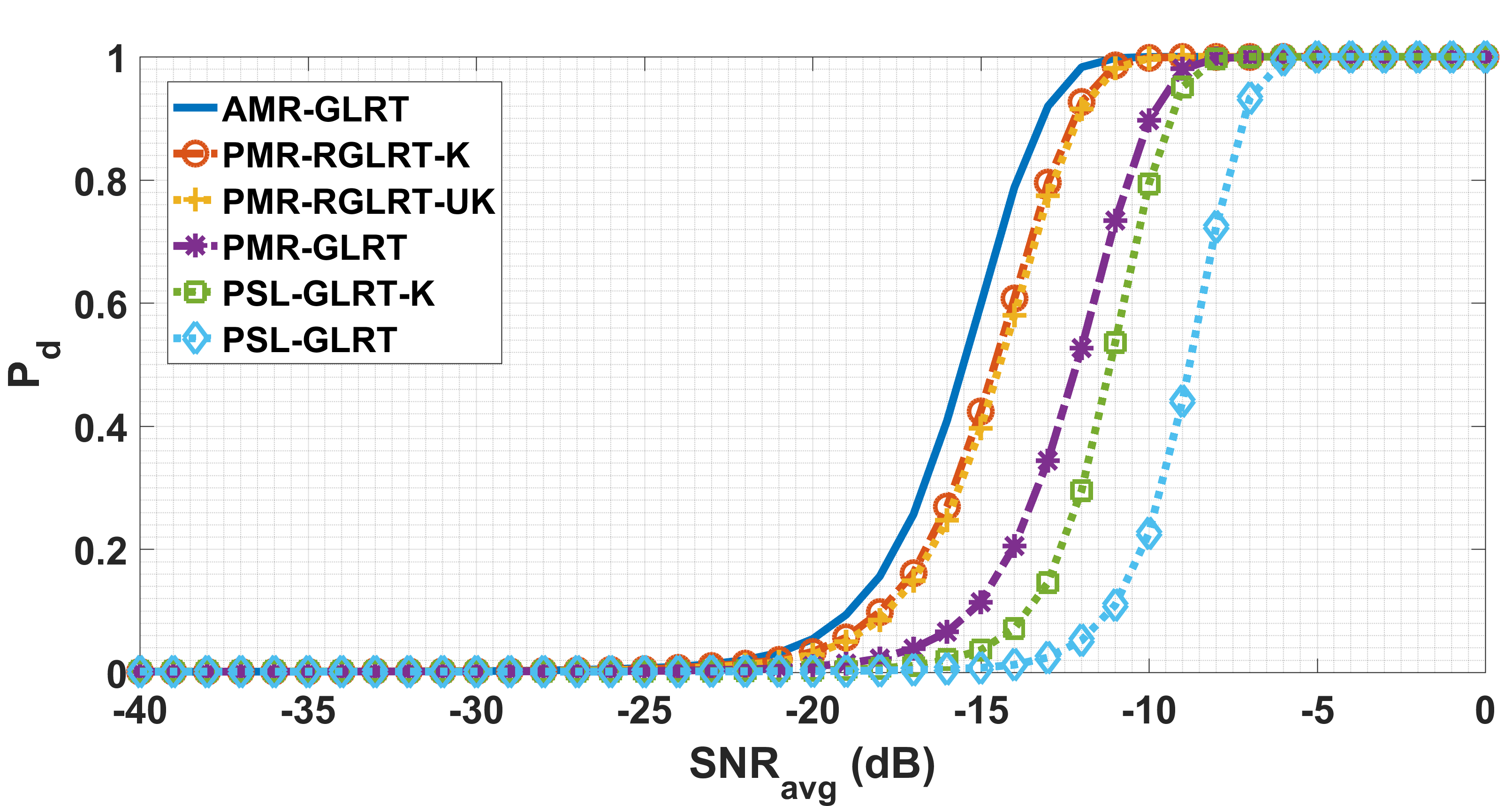}
		\caption{ }
	\end{subfigure}	
	~ 
	\begin{subfigure}[b]{\columnwidth}
		\centering
		\includegraphics[height = 2.2 in, width = \columnwidth]{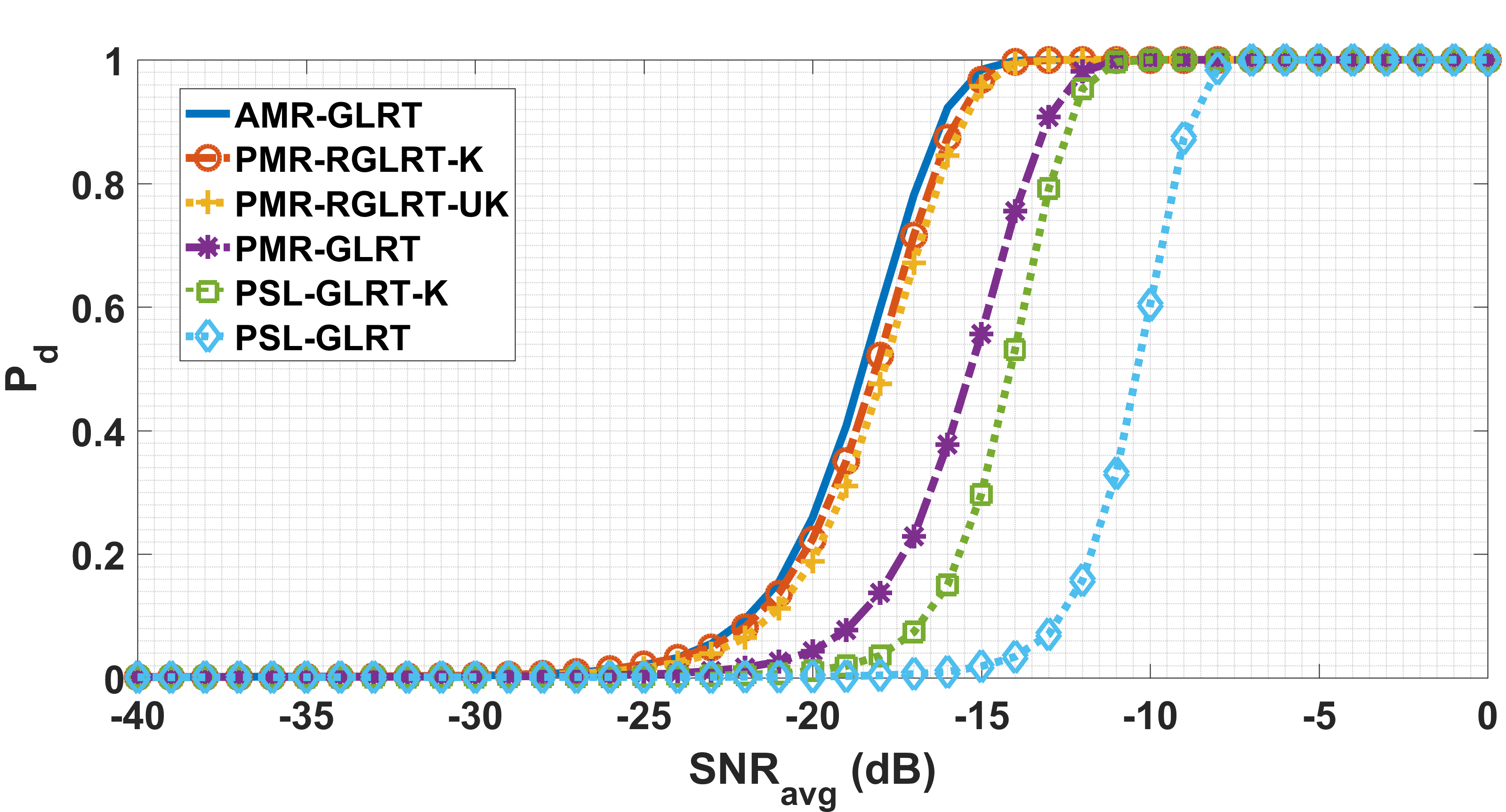}
		\caption{ }
	\end{subfigure}	
	
	~ 
	\begin{subfigure}[b]{\columnwidth}
		\centering
		\includegraphics[height = 2.2 in, width = \columnwidth]{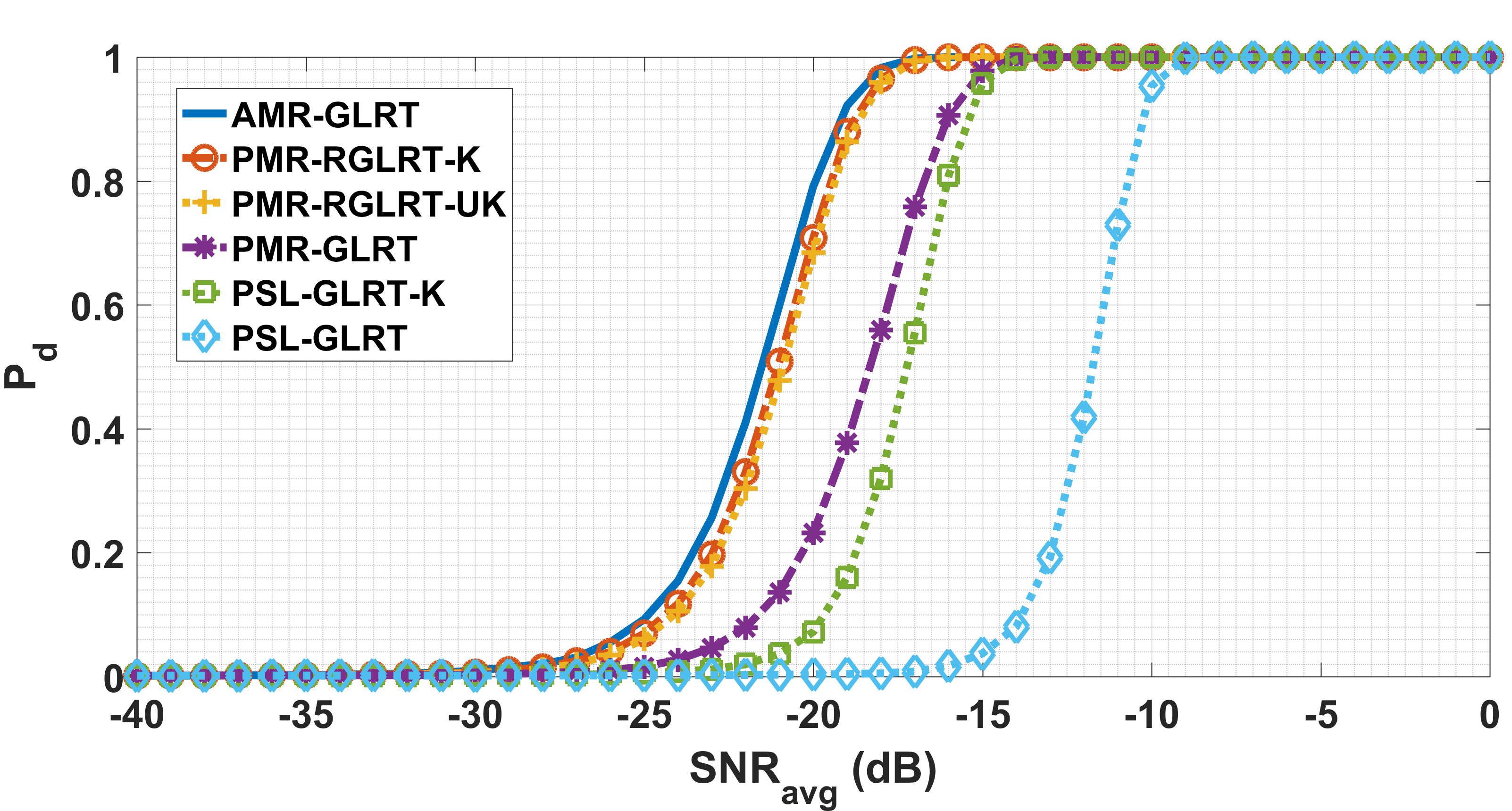}
		\caption{ }
	\end{subfigure}		
	
	\caption{$P_d$ curves as a function of $\mbox{SNR}_{avg}$ when the transmitted signal is an OFDM signal with $N_s = 16$ subcarriers and $\mbox{DNR}_{avg} = -5$ dB for different values of samples per symbol $P$, (a) $P = 4$, (b) $P = 8$, (c) $P = 16$.}\label{Pd_ofdm_results_dnr_ne5dB}
\end{figure}

Figures \ref{Pd_bpsk_results_dnr_ne10dB}--\ref{Pd_ofdm_results_dnr_ne5dB} show the $P_d$ curves as a function of $\mbox{SNR}_{avg}$ for $\mbox{DNR}_{avg} = \{-10, -5\}$ dB and for different values of samples per symbol, $P$.  As we can see from the numerical results, the proposed target detectors significantly outperform the GLRT-based target detectors that do not use the available signal format information under the considered values of $\mbox{DNR}_{avg}$ for both the PMR and PSL networks. We also see the GLRTs for target detection in the PMR networks offer better performance than the GLRT for target detection in PSL networks. This performance gain in the PMR networks is mainly due to the availability of the direct-path reference channel signals. The direct-path reference channel signals provide us some knowledge about the transmitted signal depending on the received strength of these signals \cite{Hack_PMR}.

As we see in Figures \ref{Pd_bpsk_results_dnr_ne10dB}--\ref{Pd_ofdm_results_dnr_ne5dB}, the detection performance of relaxed GLRT-based target detectors improves significantly with increasing $P$ when compared to \emph{PMR-GLRT} and \emph{PSL-GLRT}\footnote{The target detection performance of \emph{PMR-GLRT} and \emph{PSL-GLRT} improve with increasing number of samples. However, they improve at a much slower rate when compared to the proposed relaxed GLRT-based target detectors.}. This performance gain is primarily due to the lower number of parameters that need to be estimated for the GLRT in the known signal format case. For a sufficiently large value of $P$, we can also see that the performance of the proposed target detectors is close to that of an active radar, which has complete knowledge of the transmitted signal. Also, at higher values of $\mbox{DNR}_{avg}$; the proposed target detectors achieve near \emph{AMR-GLRT} level performance for smaller values of $P$. Finally, we observe no significant loss in the detection performance from not knowing noise variance in the proposed target detectors for all the considered cases.

\subsubsection{Performance comparison with unrelaxed GLRT (PMR-GLRT-K)} 
In our work, we introduced a relaxation on the complex symbols $\bm{b}^i$ to make the search for the MLE tractable. We now compare the performance of the relaxed GLRT to the exact unrelaxed GLRT to study the performance loss caused by using the relaxation. The exact GLRT which uses the signal format information is obtained by searching across all possible sequences of $\bm{b}^i$ and finding the sequence that maximizes the likelihood. The \emph{Passive MIMO Radar GLRT using the signal format information} (abbreviated as \emph{PMR-GLRT-K}) is given by 
\begin{eqnarray}
\max_{\{\bm{\mu}_s, \bm{\mu}_r, \bm{b}\} \in \mathbb{C}^{N_rN_t} \times \mathbb{C}^{N_rN_t} \times \mathbb{A}^{\mathcal{B}} } l_1(\bm{\mu}_s, \bm{\mu}_r, \bm{b}|\bm{s}) \nonumber \\
- \max_{\{ \bm{\mu}_r, \bm{b}\} \in \mathbb{C}^{N_rN_t} \times \mathbb{A}^{\mathcal{B}} } l_0( \bm{\mu}_r, \bm{b}|\bm{s}) \LRT{\mathcal{H}_1}{\mathcal{H}_0} \kappa_{pmrk}
\end{eqnarray}
where $\kappa_{pmrk}$ denotes a threshold corresponding to a desired false alarm probability and $\mathbb{A}$ is the finite set of complex symbols from which the complex symbols $\bm{b}^i$ are taken. 

For this comparison, the transmitted signal is assumed to be an OFDM signal and is generated according to (\ref{OFDM_sym}). The number of sub-carriers in the OFDM symbol is $8$, $T_g$ is $0$ $\mu s$ and BPSK symbols are modulated on each sub-carrier of the OFDM symbol. We use $1$ OFDM symbol (total of $8P$ samples) for target detection. The reference and surveillance signal samples are generated on each Monte Carlo trial according to the approach described in Section \ref{sim_setup}. The direct-path signal-to-noise ratio, $\mbox{DNR}_{avg}$, is $-10$ dB. The detection threshold that achieves a $(P_f)$ of $10^{-3}$ is determined empirically using $10^4$ trials under $\mathcal{H}_0$, and $P_d$ is estimated using $10^4$ trials under $\mathcal{H}_1$. 

\begin{figure}[t]
	\centering
	\begin{subfigure}[b]{\columnwidth}
		\centering
		\includegraphics[height = 2.2 in, width = \columnwidth]{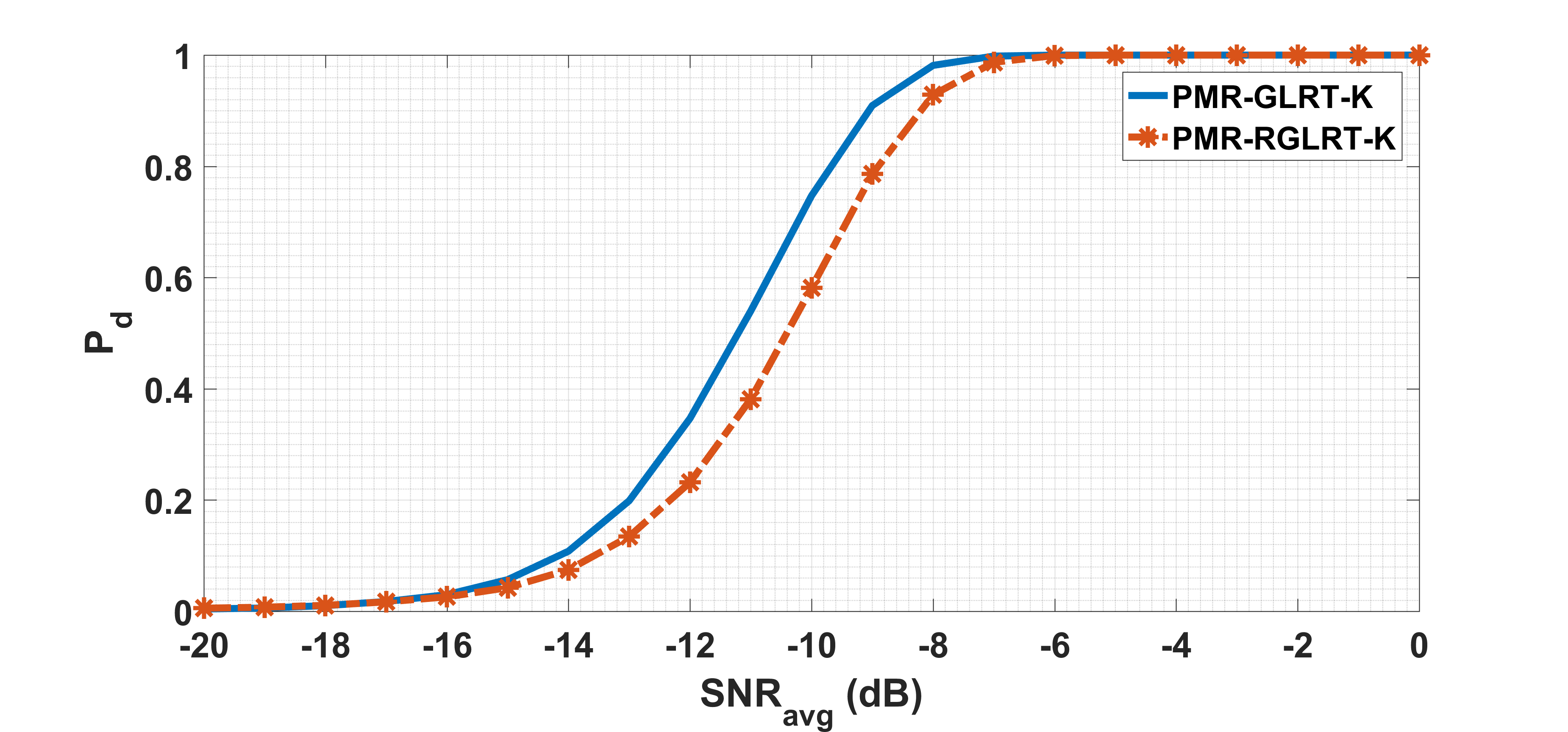}
		\caption{ }
	\end{subfigure}	
	~ 
	\begin{subfigure}[b]{\columnwidth}
		\centering
		\includegraphics[height = 2.2 in, width = \columnwidth]{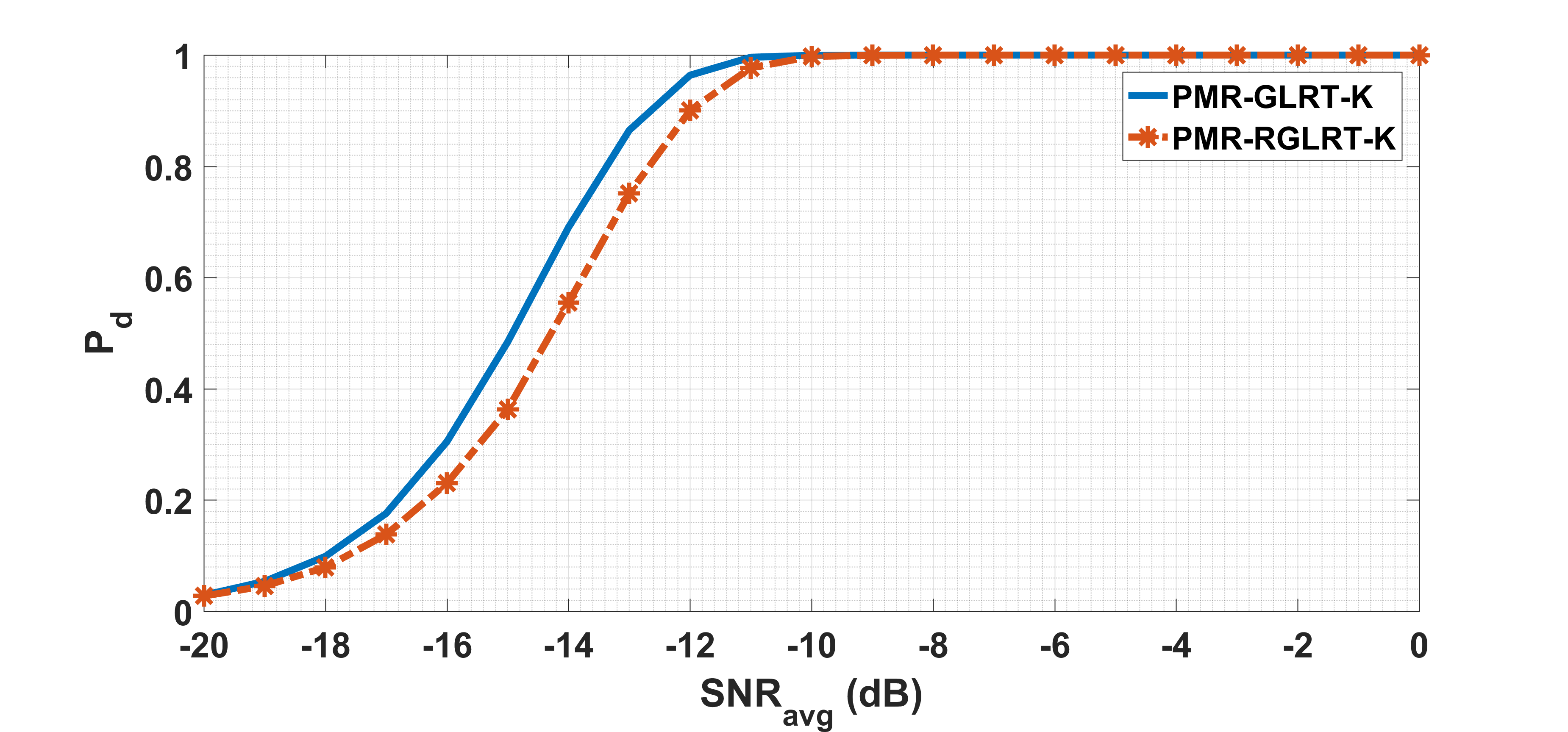}
		\caption{ }
	\end{subfigure}	
	
	~ 
	\begin{subfigure}[b]{\columnwidth}
		\centering
		\includegraphics[height = 2.2 in, width = \columnwidth]{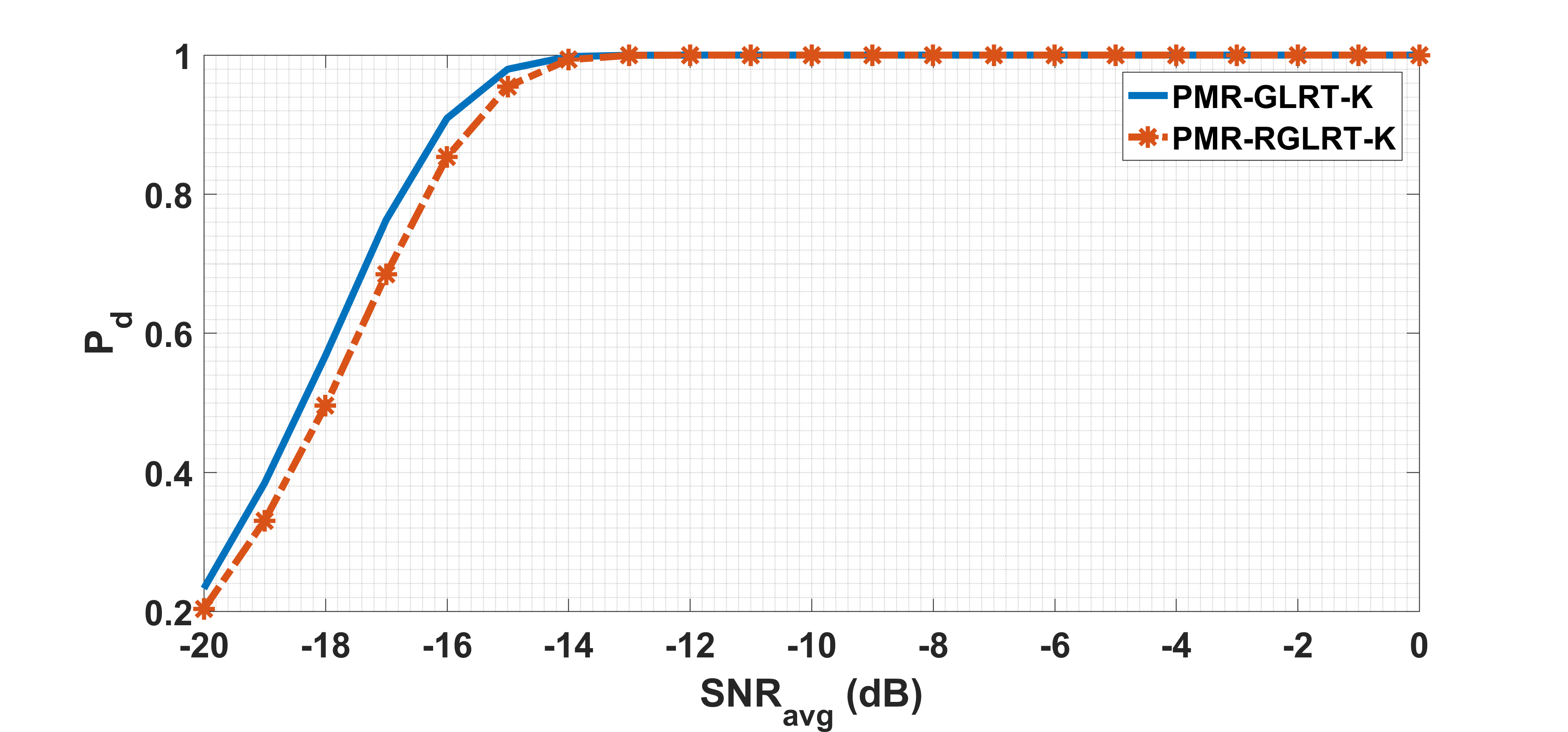}
		\caption{ }
	\end{subfigure}		
	
	\caption{$P_d$ curves as a function of $\mbox{SNR}_{avg}$ when the transmitted signal is an OFDM signal with $N_s = 8$ subcarriers and $\mbox{DNR}_{avg} = -10$ dB for different values of samples per symbol $P$, (a) $P = 4$, (b) $P = 8$, (c) $P = 16$.}\label{Pd_bpsk_comp_results_dnr_ne10dB}
\end{figure}

Since $\bm{b}^i \in \mathbb{A}^{8}$, we search across all $2^8$ possible sequences to get the MLE of $\bm{b}^i$. Figure \ref{Pd_bpsk_comp_results_dnr_ne10dB} shows us the performance loss of using the relaxation for different values of $P$. We can see from the results that the performance loss in the target detection due to the relaxation is relatively small and with increasing samples per symbol, there appears to be no performance loss by using the relaxation.

\section{Conclusion}
This work presented novel GLRT-based passive radar target detectors that can use the available signal format information under conditions where either the noise variance is known or unknown. We restrict ourselves to a particular class of transmitted signals and show that many important communication signals including CDMA, WCDMA, DVB-S, DVB-T and IEEE 802.16 WiMAX fall under the class of transmitted signals considered in this paper. As demonstrated, adding additional known information about a transmitted signal into the GLRT improves performance in comparison to a GLRT where the information is not utilized, and the signal is considered entirely unknown. Further, given an adequate number of samples per symbol, the proposed target detectors may be used to close the performance gap between the passive and active radar.

\appendix
\subsection{Derivation of PMR-RGLRT-K when the signal format information is employed}\label{pmr_glrt_with_sig_noise_info}
\noindent
Consider hypothesis $\mathcal{H}_1$ in (\ref{Hypotheses_test_linear}). We have
\begin{eqnarray}\label{LL_appendix1}
l_1(\bm{\mu}_s, \bm{\mu}_r, \bm{b}|\bm{s})  & = & \sum_{i=1}^{N_t} l_1^i(\bm{\mu}_s^i, \bm{\mu}_r^i, \bm{b}^i|\bm{s}^i), 
\end{eqnarray}
where (ignoring the additive constants), we have
\begin{flalign*}
l_1^i(\bm{\mu}_s^i, \bm{\mu}_r^i, \bm{b}^i|\bm{s}^i) &  &
\end{flalign*}
\begin{eqnarray}\label{LL1_i_appendix1}
= -\frac{1}{\sigma^2}\sum_{j=1}^{N_r} \big( ||\bm{s}_s^{ij} - \mu_s^{ij}\bm{G}^i\bm{b}^i||^2 + ||\bm{s}_r^{ij} - \mu_r^{ij}\bm{G}^i\bm{b}^i||^2 \big).
\end{eqnarray}
The MLE of $\mu_{(s,r)}^{ij}$ obtained from a derivative of (\ref{LL1_i_appendix1}) is given by
\begin{eqnarray}\label{mu_mle_appendix_1}
\hat{\mu}_{(s,r)}^{ij} & = & \frac{(\bm{G}^i\bm{b}^i)^H \bm{s}_{(s,r)}^{ij}}{(\bm{G}^i\bm{b}^i)^H \bm{G}^i\bm{b}^i}.
\end{eqnarray}
Substituting (\ref{mu_mle_appendix_1}) into (\ref{LL1_i_appendix1}), we obtain
\begin{flalign*}
l_1^i(\bm{\mu}_s^i, \bm{\mu}_r^i, \bm{b}^i|\bm{s}^i) &  &
\end{flalign*}
\begin{eqnarray}
= \frac{-1}{\sigma^2}\sum_{j=1}^{N_r} \Bigg[ ||\bm{s}_s^{ij}||^2 + ||\bm{s}_r^{ij}||^2  - \frac{(\bm{G}^i\bm{b}^i)^H\bm{s}_s^{ij} (\bm{s}_s^{ij})^H\bm{G}^i\bm{b}^i}{(\bm{G}^i\bm{b}^i)^H\bm{G}^i\bm{b}^i} \nonumber \\
- \frac{(\bm{G}^i\bm{b}^i)^H\bm{s}_r^{ij} (\bm{s}_r^{ij})^H\bm{G}^i\bm{b}^i}{(\bm{G}^i\bm{b}^i)^H\bm{G}^i\bm{b}^i} \Bigg].
\end{eqnarray}
After simplifying, we obtain
\begin{eqnarray}\label{MaximEq_appendix1}
l_1^i(\hat{\bm{\mu}}_s^i, \hat{\bm{\mu}}_r^i, \bm{b}^i|\bm{s}^i) = \frac{-1}{\sigma^2}\Bigg[ E_{sr}^i -  \frac{(\bm{G}^i\bm{b}^i)^H\bm{\phi}_1^i (\bm{\phi}_1^i)^H\bm{G}^i\bm{b}^i}{(\bm{G}^i\bm{b}^i)^H\bm{G}^i\bm{b}^i} \Bigg],  
\end{eqnarray}
where $\bm{\phi}_1^i = [\bm{\phi}_s^i, \bm{\phi}_r^i]$, and the matrices $\bm{\phi}_s^i$ and $\bm{\phi}_r^i$ are defined as
\begin{eqnarray}
\bm{\phi}^i_{(s, r)} & = & \left[ \bm{s}_{(s, r)}^{i1}, \bm{s}_{(s, r)}^{i2}, \cdots, \bm{s}_{(s, r)}^{iN_r} \right] \nonumber
\end{eqnarray}
and the scalar $E_{sr}^i = ||\bm{s}_s^i||^2 + ||\bm{s}_r^i||^2$. Using the discussion below (\ref{Generalized_RayleighQuotient}), the complex value of $\bm{b}^i$ that maximizes (\ref{MaximEq_appendix1}) is given by $\hat{\bm{b}}^i = \bm{v}_1\left((\bm{G}^i)^H\bm{\phi}_1^i (\bm{\phi}_1^i)^H \bm{G}^i,  (\bm{G}^i)^H  \bm{G}^i \right)$. Substituting $\hat{\bm{b}}^i$ in (\ref{MaximEq_appendix1}), we have
\begin{flalign*}
l_1^i(\hat{\bm{\mu}}_s^i, \hat{\bm{\mu}}_r^i, \hat{\bm{b}}^i|\bm{s}^i) & = &
\end{flalign*}
\begin{eqnarray}
\frac{-1}{\sigma^2}\left[E_{sr}^i - \lambda_1\left((\bm{G}^i)^H\bm{\phi}_1^i (\bm{\phi}_1^i)^H \bm{G}^i,  (\bm{G}^i)^H \bm{G}^i \right)\right]. \nonumber
\end{eqnarray}
From (\ref{LL_appendix1}), we then have
\begin{flalign*}
l_1(\hat{\bm{\mu}}_s, \hat{\bm{\mu}}_r, \hat{\bm{b}}|\bm{s})& = &
\end{flalign*}
\begin{eqnarray}\label{GLRT_H1_pmr_knownsignal}
\frac{-1}{\sigma^2}\sum_{i=1}^{N_t} \left(E_{sr}^i - \lambda_1\left((\bm{G}^i)^H \bm{\phi}_1^i (\bm{\phi}_1^i)^H \bm{G}^i,  (\bm{G}^i)^H \bm{G}^i \right)\right). 
\end{eqnarray}
Following a similar procedure, it can be shown under $\mathcal{H}_0$ that
\begin{flalign*}
l_0(\hat{\bm{\mu}}_r, \hat{\bm{b}}|\bm{s})& = &
\end{flalign*}
\begin{eqnarray}\label{GLRT_H1_pmr_knownsignal}
\frac{-1}{\sigma^2}\sum_{i=1}^{N_t} \left(E_{sr}^i - \lambda_1\left((\bm{G}^i)^H \bm{\phi}_r^i (\bm{\phi}_r^i)^H \bm{G}^i,  (\bm{G}^i)^H \bm{G}^i \right)\right). 
\end{eqnarray}
Using $l_1(\hat{\bm{\mu}}_s, \hat{\bm{\mu}}_r, \hat{\bm{b}}|\bm{s})$ and $l_0(\hat{\bm{\mu}}_r, \hat{\bm{b}}|\bm{s})$, the \emph{PMR-RGLRT-K} for the hypothesis testing problem in (\ref{Hypotheses_test_linear}) is given by
\begin{eqnarray}
\xi_{ksf} & = &\frac{1}{\sigma^2}\sum_{i=1}^{N_t} \Big[ \lambda_1\left((\bm{G}^i)^H\bm{\phi}_1^i (\bm{\phi}_1^i)^H\bm{G}^i,  (\bm{G}^i)^H \bm{G}^i \right) \nonumber \\
&&- \lambda_1\left((\bm{G}^i)^H\bm{\phi}_r^i (\bm{\phi}_r^i)^H \bm{G}^i,  (\bm{G}^i)^H \bm{G}^i \right) \Big] \nonumber \\
&&\LRT{\mathcal{H}_1}{\mathcal{H}_0} \kappa_{ksf}. 
\end{eqnarray}
\subsection{Derivation of PSL-RGLRT-K when the signal format information is employed}\label{psl_glrt_sig_noise_info}
\noindent
The conditional probability density function (pdf) of $\bm{s}$ under $\mathcal{H}_1$ for the hypotheses test of (\ref{Hypotheses_test_PSL}) is given by
\begin{eqnarray}
p_1(\bm{s}|\bm{\mu}_s, \bm{b}) & = & \prod_{i=1}^{N_t} p_1^i (\bm{s}^i| \bm{\mu}_s^i, \bm{b}^i),
\end{eqnarray}
where
\begin{eqnarray}
p_1^i (\bm{s}^i| \bm{\mu}_s^i, \bm{b}^i)  & \propto & \exp\bigg\{\frac{-1}{\sigma^2}\sum_{j=1}^{N_r}  ||\bm{s}_s^{ij} - \mu_s^{ij}\bm{G}^i\bm{b}^i||^2 \bigg\}.
\end{eqnarray}
The conditional pdf of $\bm{s}$ under $\mathcal{H}_0$, $p_0(\bm{s})$, is similarly defined. Let $l_1(\bm{\mu}_s,  \bm{b}|\bm{s}) = \log p_1(\bm{s}|\bm{\mu}_s, \bm{b})$ and $l_0(\bm{s}) = \log p_0(\bm{s})$ denote the log-likelihood functions under the hypotheses $\mathcal{H}_1$ and $\mathcal{H}_0$. The relaxed GLRT can now be written as
\begin{eqnarray}
\max_{\{\bm{\mu}_s, \bm{b}\} \in \mathbb{C}^{N_rN_t} \times \mathbb{C}^{\mathcal{B}} } l_1(\bm{\mu}_s, \bm{b}|\bm{s}) -  l_0( \bm{s}) \LRT{\mathcal{H}_1}{\mathcal{H}_0} \kappa_{psk}.
\end{eqnarray}
Consider hypothesis $\mathcal{H}_1$. We have
\begin{eqnarray}\label{LL_appendix3}
l_1(\bm{\mu}_s, \bm{b}|\bm{s})  & = & \sum_{i=1}^{N_t} l_1^i(\bm{\mu}_s^i, \bm{b}^i|\bm{s}^i), 
\end{eqnarray}
where (ignoring the additive constants), we have
\begin{eqnarray}\label{LL1_i_appendix3}
l_1^i(\bm{\mu}_s^i, \bm{b}^i|\bm{s}^i) & = & \frac{-1}{\sigma^2}\sum_{j=1}^{N_r}  ||\bm{s}_s^{ij} - \mu_s^{ij}\bm{G}^i\bm{b}^i||^2.
\end{eqnarray}
The relaxed MLE of $\mu_{s}^{ij}$ and $\hat{\bm{b}}^i$ are obtained from a derivative of (\ref{LL1_i_appendix3}) and are given by
\begin{eqnarray}\label{mu_mle_appendix1}
\hat{\mu}_{s}^{ij} & = & \frac{(\bm{G}^i\bm{b}^i)^H \bm{s}_{s}^{ij}}{(\bm{G}^i\bm{b}^i)^H \bm{G}^i\bm{b}^i},
\end{eqnarray}
and
\begin{eqnarray}
\hat{\bm{b}}^i & = & \bm{v}_1\left((\bm{G}^i)^H\bm{\phi}_s^i (\bm{\phi}_s^i)^H\bm{G}^i,  (\bm{G}^i)^H \bm{G}^i \right).
\end{eqnarray}
Substituting the obtained relaxed MLE in (\ref{LL1_i_appendix3}) and simplifying, we obtain
\begin{eqnarray}
l_1^i(\hat{\bm{\mu}}_s^i, \hat{\bm{b}}^i|\bm{s}^i) = \frac{-1}{\sigma^2}\left[E_{sr}^i - \lambda_1\left((\bm{G}^i)^H\bm{\phi}_s^i (\bm{\phi}_s^i)^H \bm{G}^i,  (\bm{G}^i)^H \bm{G}^i \right)\right]. \nonumber
\end{eqnarray}
From (\ref{LL_appendix3}), we then have
\begin{flalign*}
l_1(\hat{\bm{\mu}}_s, \hat{\bm{\mu}}_r, \hat{\bm{b}}|\bm{s})& = &
\end{flalign*}
\begin{eqnarray}\label{GLRT_H1_psl_knownsignal}
\frac{-1}{\sigma^2}\sum_{i=1}^{N_t} \left(E_{sr}^i - \lambda_1\left((\bm{G}^i)^H \bm{\phi}_s^i (\bm{\phi}_s^i)^H \bm{G}^i,  (\bm{G}^i)^H \bm{G}^i \right)\right). 
\end{eqnarray}
By a similar procedure, it can shown under $\mathcal{H}_0$ that
\begin{eqnarray}\label{GLRT_H0_psl_knownsignal}
l_0(\bm{s})& = & -\frac{1}{\sigma^2}\sum_{i=1}^{N_t} E_{sr}^i
\end{eqnarray}
Using $l_1(\hat{\bm{\mu}}_s, \hat{\bm{b}}|\bm{s})$ and $l_0(\bm{s})$, the \emph{PSL-RGLRT-K} for the hypothesis testing problem in (\ref{Hypotheses_test_PSL}) is given by
\begin{eqnarray}
\xi_{psk} = \frac{1}{\sigma^2}\sum_{i=1}^{N_t} \lambda_1\left((\bm{G}^i)^H \bm{\phi}_s^i (\bm{\phi}_s^i)^H \bm{G}^i,  (\bm{G}^i)^H \bm{G}^i \right) \nonumber \\
\LRT{\mathcal{H}_1}{\mathcal{H}_0} \kappa_{psk}. 
\end{eqnarray}
\subsection{Derivation of PMR-RGLRT-UK when the signal format information is employed}\label{pmr_glrt_with_sig_info}
\noindent
Consider hypothesis $\mathcal{H}_1$ in (\ref{Hypotheses_test_linear}). We have
\begin{flalign*}
l_1(\bm{\mu}_s, \bm{\mu}_r, \bm{b}, \sigma^2|\bm{s}) & = & 
\end{flalign*}
\begin{eqnarray}\label{LLu1}
&  & -N_tN_rN\ln(\pi\sigma^2)  -\frac{1}{\sigma^2}\sum_{i=1}^{N_t}\sum_{j=1}^{N_r} \big( ||\bm{s}_s^{ij} - \mu_s^{ij}\bm{G}^i\bm{b}^i||^2 \nonumber \\
& &  + ||\bm{s}_r^{ij} - \mu_r^{ij}\bm{G}^i\bm{b}^i||^2  \big).
\end{eqnarray}
From Appendix \ref{pmr_glrt_with_sig_noise_info}, the relaxed MLE of $\mu_{(s,r)}^{ij}$ and $\bm{b}^i$ are given by
\begin{eqnarray}
\hat{\mu}_{(s,r)}^{ij} & = & \frac{(\bm{G}^i\bm{b}^i)^H\bm{s}_{(s,r)}^{ij}}{||\bm{G}^i\bm{b}^i||^2},
\end{eqnarray}
and
\begin{eqnarray}
\hat{\bm{b}}^i & = & \bm{v}_1\left((\bm{G}^i)^H\bm{\phi}_1^i (\bm{\phi}_1^i)^H\bm{G}^i,  (\bm{G}^i)^H\bm{G}^i \right).
\end{eqnarray}
Substituting these values in (\ref{LLu1}), we obtain
\begin{flalign*}
l_1(\hat{\bm{\mu}}_s, \hat{\bm{\mu}}_r, \hat{\bm{b}}, \sigma^2|\bm{s}) & = -N_tN_rN\ln(\pi\sigma^2)& 
\end{flalign*}
\begin{eqnarray}\label{LL_glrt_u}
-\frac{1}{\sigma^2}\sum_{i=1}^{N_t} \Big[ E^i - \lambda_1\left((\bm{G}^i)^H\bm{\phi}_1^i (\bm{\phi}_1^i)^H\bm{G}^i,  (\bm{G}^i)^H\bm{G}^i \right) \Big]. 
\end{eqnarray}
The MLE of $\sigma^2$, denoted by $\hat{\sigma}^2$, can be obtained from the derivate of (\ref{LL_glrt_u}) and is given by
\begin{eqnarray}
\hat{\sigma}^2 = \frac{1}{c_1}\sum_{i=1}^{N_t} \Big[ E^i - \lambda_1\left((\bm{G}^i)^H\bm{\phi}_1^i (\bm{\phi}_1^i)^H\bm{G}^i,  (\bm{G}^i)^H\bm{G}^i \right) \Big]. 
\end{eqnarray}
where $c_1 = N_tN_rN$. Substituting the obtained MLE in $l_1(\hat{\bm{\mu}}_s, \hat{\bm{\mu}}_r, \hat{\bm{b}}, \sigma^2|\bm{s})$ and simplifying, we have (ignoring the additive constant)
\begin{flalign*}
l_1(\hat{\bm{\mu}}_s, \hat{\bm{\mu}}_r, \hat{\bm{b}}, \hat{\sigma}^2|\bm{s}) &=  &
\end{flalign*}
\begin{eqnarray}\label{GLRTU_H1}
-c_1\ln \Bigg(\sum_{i=1}^{N_t} \Big[ E^i - \lambda_1\left((\bm{G}^i)^H\bm{\phi}_1^i (\bm{\phi}_1^i)^H\bm{G}^i,  (\bm{G}^i)^H\bm{G}^i \right) \Big] \Bigg). \nonumber
\end{eqnarray}
By a similar procedure, it can shown under hypotheses $\mathcal{H}_0$ that
\begin{flalign*}
l_0(\hat{\bm{\mu}}_r, \hat{\bm{b}}, \hat{\sigma}^2|\bm{s}) & = &
\end{flalign*}
\begin{eqnarray}\label{GLRTU_H0}
-c_1\ln \Bigg(\sum_{i=1}^{N_t} \Big[ E^i - \lambda_1\left((\bm{G}^i)^H\bm{\phi}_r^i (\bm{\phi}_r^i)^H\bm{G}^i,  (\bm{G}^i)^H\bm{G}^i \right) \Big] \Bigg). \nonumber 
\end{eqnarray}
Using $l_1(\hat{\bm{\mu}}_s, \hat{\bm{\mu}}_r, \hat{\bm{b}}, \hat{\sigma}^2|\bm{s})$ and $l_0(\hat{\bm{\mu}}_r, \hat{\bm{b}}, \hat{\sigma}^2|\bm{s})$, the \emph{PMR-RGLRT-UK} for the hypothesis testing problem in (\ref{Hypotheses_test_linear}) is given by
\begin{eqnarray}
\xi_{uk} & = & \frac{\sum_{i=1}^{N_t} \Big[ E^i - \lambda_1\left((\bm{G}^i)^H\bm{\phi}_r^i (\bm{\phi}_r^i)^H\bm{G}^i,  (\bm{G}^i)^H\bm{G}^i \right) \Big]}{\sum_{i=1}^{N_t} \Big[ E^i - \lambda_1\left((\bm{G}^i)^H\bm{\phi}_1^i (\bm{\phi}_1^i)^H\bm{G}^i,  (\bm{G}^i)^H\bm{G}^i \right) \Big]} \nonumber \\
& & \LRT{\mathcal{H}_1}{\mathcal{H}_0} \kappa_{uk}.
\end{eqnarray}

\bibliographystyle{IEEEtran}
\bibliography{references}

\end{document}